\documentclass[fleqn,usenatbib]{mnras}
\usepackage[T1]{fontenc}
\usepackage{color}

\definecolor{olive}{RGB}{48,188,0}

\DeclareRobustCommand{\VAN}[3]{#2}
\let\VANthebibliography\thebibliography
\def\thebibliography{\DeclareRobustCommand{\VAN}[3]{##3}\VANthebibliography}

\usepackage{graphicx}	
\usepackage{amsmath}	
\usepackage{amssymb}	
\usepackage{subfig}
\usepackage{gensymb}
\usepackage{tabularx}

\usepackage{threeparttable}

\usepackage{newtxtext,newtxmath}

\title[Gas homogeneity and turbulence in the Perseus Cluster]{\textit{Chandra} measurements of gas homogeneity and turbulence at intermediate radii in the Perseus Cluster}

\author[Martijn de Vries et al.]{%
Martijn de Vries$^{1}$\thanks{E-mail: mndvries@stanford.edu},
Adam B. Mantz$^{1}$,
Steven W. Allen$^{1,2,3}$,
R. Glenn Morris$^{1,2}$,
Irina Zhuravleva$^{4}$, 
\newauthor Rebecca E. A. Canning$^{5}$,
Steven R. Ehlert$^{6}$, 
Anna Ogorza\l ek$^{7,8}$,
Aurora Simionescu$^{9,10,11}$
and Norbert Werner$^{12}$
\\
$^{1}$ Kavli Institute for Particle Astrophysics and Cosmology, Stanford University, 452 Lomita Mall, Stanford, CA 94305, USA \\
$^{2}$ SLAC National Accelerator Laboratory, 2575 Sand Hill Road, Menlo Park, CA 94025, USA \\
$^{3}$ Department of Physics, Stanford University, 382 Via Pueblo Mall, Stanford, CA 94305, USA \\
$^{4}$ Department of Astronomy and Astrophysics, University of Chicago, Chicago IL 60637, USA \\
$^{5}$ Institute of Cosmology and Gravitation, University of Portsmouth, Portsmouth, PO1 3FX, UK \\
$^{6}$ NASA Marshall Space Flight Center, Huntsville, AL 35812, USA \\
$^{7}$ NASA Goddard Space Flight Center, Code 662, Greenbelt, MD 20771, USA\\
$^{8}$ Department of Astronomy, University of Maryland, College Park MD 20742-2421, USA \\
$^{9}$ SRON Netherlands Institute for Space Research, Niels Bohrweg 4, 2333 CA Leiden, The Netherlands \\
$^{10}$ Leiden Observatory, Leiden University, PO Box 9513, NL-2300 RA Leiden, The Netherlands \\
$^{11}$ Kavli Institute for the Physics and Mathematics of the Universe, University of Tokyo, Kashiwa 277-8583, Japan \\
$^{12}$ Department of Theoretical Physics and Astrophysics, Faculty of Science, Masaryk University, Brno, Czech Republic \\
}

\date{Accepted XXX. Received YYY; in original form ZZZ}

\pubyear{2021}

\begin{document}

\label{firstpage}
\pagerange{\pageref{firstpage}--\pageref{lastpage}}
\maketitle

\begin{abstract}
We present a \textit{Chandra} study of surface brightness fluctuations in the diffuse intracluster medium of the Perseus Cluster. Our study utilizes deep, archival imaging of the cluster core as well as a new mosaic of 29 short $\,5\,$ks observations extending in 8 different directions out to radii of $r_{500} \sim 2.2\,r_{2500}$. Under the assumption that the distribution of densities at a given radius is log-normally distributed, two important quantities can be derived from the width of the log-normal density distribution on a given spatial scale: the density bias, which is equal to the square root of the clumping factor $C$; and the one-component turbulent velocity, $v_{k, 1D}$. We forward-model all contributions to the measured surface brightness, including astrophysical and particle background components, and account for the Poisson nature of the measured signal. Measuring the distribution of surface brightness fluctuations in 1 arcmin$^{2}$ regions, spanning the radial range $0.3-2.2\,r_{2500}$ ($7.8-57.3\,$arcmin), we find a small to moderate average density bias of around $3\%$ at radii below $1.6\,r_{2500}$. We also infer an average turbulent velocity at these radii of $V_{1D} <400$ km\,s$^{-1}$. Direct confirmation of our results on turbulent velocities inferred from surface brightness fluctuations should be possible using the X-ray calorimeter spectrometers to be flown aboard the \textit{XRISM} and \textit{Athena} observatories. 

\end{abstract}




\section{Introduction} 
\label{sec:intro}

The hot, diffuse, intracluster medium (ICM) is an important tool through which we can study dynamical processes in galaxy clusters. Motions within clusters on parsec to Megaparsec scales, driven by sources such as AGN feedback, galaxy motions and subcluster mergers, all act to perturb the ICM, leading to fluctuations in the density of the gas and accordingly in the observed surface brightness. Understanding the nature of these fluctuations can shed light on the microphysical properties of the ICM and the process of virialization, and in principle enable improved measurements of the gas mass and total mass of these systems, which are important for cosmological work.  

Power spectra have previously proven to be a successful method of studying cluster gas fluctuations, particularly in the cool cores of clusters. It was shown by \cite{Gaspari2014} and \cite{Zhuravleva2014} that in relaxed clusters the density fluctuations at a given length scale can be directly related to the one-component turbulent velocity at that length scale. This relation provides a powerful way to link the observable quantity of fluctuations in surface brightness to the dynamical properties of the ICM. By measuring the turbulent velocity as a function of length scale, one can
study the physical processes sourcing gas motions, and the resulting turbulent cascade from turbulent motions cascading down from larger to smaller scales, eventually converting kinetic energy into heat. It has been shown in some clusters that the turbulent dissipation of energy provides sufficient heat to balance radiative cooling, providing an important piece of the puzzle linking feedback from AGN to their host environments \citep[e.g.][]{Zhuravleva2018, Hitomi2018, Liu2021}.
Furthermore, because the X-ray emissivity depends on the temperature-dependent cooling function $\Lambda (T)$, by studying the fluctuations in different energy bands, information about the thermodynamic processes that source the fluctuations (isothermal, isobaric, or adiabatic) can be inferred \citep{Arevalo2016, Churazov2016, Zhuravleva2018}. 

The microphysical properties of the ICM remain relatively poorly understood, yet are of fundamental importance to building better models of the growth and evolution of galaxy clusters. The measurement of ICM motions provides both a new window onto these processes, and a powerful tracer of recent dynamical activity in these systems \citep{Simionescu2019}. 

A second reason to study fluctuations in the ICM is because homogeneity is often implicitly assumed when making measurements of the overall gas properties in the cluster. Because the X-ray emissivity of a thermal plasma is proportional to the density squared $\rho^2$, overdensities in the plasma will have an outsized contribution to the total flux received from a source. The result is that an inhomogeneous medium will bias measurements of the gas properties, such as density, pressure, and entropy. The inhomogeneity can be characterised through the \textit{clumping factor}:
\begin{equation}
    \label{eq:clumpf}
    C = \langle \rho ^2 \rangle/\langle \rho \rangle^2,
\end{equation} 
where $\langle \rangle$ denotes the average within a given region. It follows that $\sqrt{C}$ represents, to first order, the bias in the measured density. 

The clumping factor has been found in simulations to increase with radius \citep[e.g.][]{Zhuravleva2013, Roncarelli2013, Angelinelli2021}. Such simulations typically show that $\sqrt{C} < 1.2$ at radii below $\approx r_{500}$. Measurements of the clumping factor from observations have been made by e.g. \cite{Eckert2015} and \cite{Mirakhor2021}, by comparing the mean and median surface brightnesses within annuli at fixed cluster-centric radius. 

These studies have generally inferred values for $\sqrt{C}$ that agree with simulations, although systematic uncertainties remain: at larger radii, where the ICM emission is faint, it is necessary to have a precise understanding of all sources of background emission. As is noted above, precise measurements of the gas masses and total masses of clusters are also important for a range of techniques used to probe cosmology with galaxy clusters \citep[for a review, see][]{Allen2011}. In particular, cosmological constraints based on measurements of the gas mass fraction in clusters \citep[][and references therein]{Allen2003,Allen2004,Allen2008, Mantz2014, Mantz2022} and combinations of X-ray and Sunyaev-Zel'dovich effect measurements \citep{Mantz2014,Mantz2022, Wan2021} are directly impacted by systematic uncertainties in the clumping factor as a function of radius.  

In this work, we report on a new method of measuring fluctuations in the ICM, using a forward-modeling approach that takes into account the projected cluster emission as well as significant background components. Our method is particularly well-suited for measurements at the intermediate-to-large radii vital for cosmological studies, where accounting for the amplitudes of and uncertainties in all relevant background signals is important, and where fewer number of counts are typically available. We apply our method to a large \textit{Chandra} ACIS data set of the Perseus Cluster between $0.3$--$2.2\,r_{2500}$, using the value of $r_{2500} = 26.05 \arcmin$. At the adopted redshift $z=0.01790$ and adopting a  $\rm \Lambda CDM$ cosmology with $h=0.7$, $\Omega_M=0.3$ and $\Omega_\Lambda=0.7$, this corresponds to a a length of $564\,$kpc. By carefully modeling the projected cluster emission as a function of radius and azimuth, we aim to 1)
quantify the density bias, $\sqrt{C}$, as a function of radius along eight independent arms, and 2) assuming that these fluctuations are sourced by turbulent motions, infer the one-dimensional turbulent velocity profile of the ICM \footnote{Given this assumption, the terms `gas motions' and `turbulence velocity' are used interchangeable throughout this paper}. In order to compare our results against future observations by XRISM, we match our measurements to the expected spatial resolution of the XRISM Resolve calorimeter: $\sim 1\arcmin$. In principle, however, our method can be applied to a broad range of length scales.

This paper is structured as follows: in Section \ref{sec:data} we show the \textit{Chandra} data set of the Perseus Cluster used in this analysis, in Section \ref{sec:model} we describe the model components, the spatial layout of the model and the statistical implementation. We report on the results of the modeling in Section \ref{sec:results}, discuss sources of uncertainty and potential extensions of the method in Section \ref{sec:discussion}, and conclude in section \ref{sec:conclusion}.

\section{Observations and Data reduction}
\label{sec:data}
\begin{table}

\setlength{\tabcolsep}{3.8pt}
\caption{Overview of the \textit{Chandra} observations of the Perseus Cluster used in this paper. The `Aim' column shows whether the observation aimpoint was on the ACIS-I or ACIS-S array. The Exp column lists the exposure times in kiloseconds, after filtering out periods of high background.a}
\label{tab:obsids}

\begin{tabularx}{\linewidth}{c  c  c  c | c  c  c  c}

\hline \hline
Obs & Date & Aim & Exp & Obs & Date & Aim & Exp \\ \hline
3209 & 2002-08-08 & S & 95.8 & 17259 & 2015-12-03 & I & 4.7 \\
3237 & 2003-03-15 & S & 93.9 & 17260 & 2015-12-01 & I & 5.0 \\
4289 & 2002-08-10 & S & 95.4 &  17261 & 2015-12-01 & I & 5.0\\
4946 & 2004-10-06 & S & 23.7 & 17262 & 2015-12-07 & I & 4.7 \\
4947 & 2004-10-11 & S & 29.8 &  17263 & 2015-12-04 & I & 4.7 \\
4948 & 2004-10-09 & S & 118.6 & 17264 & 2015-12-01 & I & 4.7 \\
4949 & 2004-10-12 & S & 29.4 & 17265 & 2015-12-07 & I & 5.0 \\
4950 & 2004-10-12 & S & 96.9 & 17266 & 2015-12-04 & I & 4.7  \\
4951 & 2004-10-17 & S & 96.1 & 17267 & 2015-12-12 & I & 5.0 \\
4952 & 2004-10-14 & S & 164.2 & 17268 & 2015-12-01 & I & 4.7\\
4953 & 2004-10-18 & S & 30.1 &  17269 & 2015-12-01 & I & 4.7  \\
5597 & 2004-12-23 & I & 25.2 & 17270 & 2015-12-12 & I & 4.7 \\
6139 & 2004-10-04 & S & 56.4 & 17271 & 2015-12-10 & I & 5.0 \\
6145 & 2004-10-19 & S & 85.0 & 17272 & 2015-12-07 & I & 5.0 \\
6146 & 2004-10-20 & S & 47.1 & 17273 & 2015-12-11 & I & 5.0 \\
8473 & 2006-11-14 & S & 29.7 &  17274 & 2015-12-09 & I & 5.0  \\
11713 & 2009-11-29 & I & 112.2 & 17275 & 2015-12-09 & I & 4.7\\
11714 & 2009-12-07 & I & 92.0 &  17276 & 2015-12-09 & I & 5.0\\
11715 & 2009-12-02 & I & 73.4 &  17277 & 2015-12-10 & I & 4.7 \\
11716 & 2009-10-10 & I & 39.6 & 17278 & 2015-12-10 & I & 4.7 \\
12025 & 2009-11-25 & I & 17.9 & 17279 & 2015-11-30 & I & 4.7   \\
12033 & 2009-11-27 & I & 18.9 & 17280 & 2015-12-11 & I & 4.7 \\ 
12036 & 2009-12-02 & I & 47.9 & 17281 & 2015-12-11 & I & 4.7 \\
12037 & 2009-12-05 & I & 84.6 & 17282 & 2015-12-11 & I & 4.7 \\
13989 & 2011-11-07 & I & 38.2 & 17283 & 2015-10-06 & I & 5.0 \\
13990 & 2011-11-11 & I & 37.1 & 17284 & 2015-10-06 & I & 4.7 \\
13991 & 2011-11-05 & I & 37.1 & 17285 & 2015-10-06 & I & 4.7 \\
13992 & 2011-11-05 & I & 36.8 &  17286 & 2015-10-06 & I & 4.7\\
17258 & 2015-12-03 & I & 5.0 & \\ \hline
\end{tabularx}

\end{table}

This paper uses a large set of \textit{Chandra} observations of the Perseus cluster. Most important to the present work are a set of observations with average exposure times of $\sim 5$ ks, which extend outward from the cool core to $\sim 2.5\,r_{2500}$ along 8 arms and together with archival data provide nearly complete coverage of the field until $\sim 1.2\,r_{2500}$. The full list of observations is given in Table \ref{tab:obsids}. Because our modeling relies on accurate finding and masking of point sources, we limit ourselves to the four chips on the ACIS-I array for observations in ACIS-I mode, and the ACIS-S3 chip for observations in ACIS-S mode.

We used the same reprocessed data set as in \cite{Mantz2022}, using \textsc{CIAO} 4.9 and \textsc{CALDB} 4.7.6. We reduced the data following the procedures described in \cite{Mantz2014} and \cite{Mantz2015}, screening the observations for periods of high background and filtering out those time intervals. We also created exposure maps for each chip using the available \textsc{CIAO} tools, using a representative cluster emission spectrum (a thermal plasma with $kT = 7$\,keV, and Galactic absorption $n_H=1.36\times10^{21}$cm$^{-2}$) as the energy weighting. All images created use an energy range of $0.6$--$3.5$ keV (see also Section \ref{sec:mod:eqs}).

We utilized the ACIS `stowed' backgrounds to create particle background maps. For each ACIS chip, the corresponding stowed background was retrieved from the calibration database. For each chip, we define an exposure scaling factor by scaling the stowed backgrounds to the number of counts between 9.5 and 12 keV. The appropriate exposure scaling is handled within the model (see Section \ref{sec:mod:pb} for more information), so that the Poisson errors are correctly accounted for and the uncertainty in the energy scaling can be marginalized over. 

For each ObsID, we identified point sources using the Cluster AGN Topography Survey pipeline (CATS: Canning et al. in prep). We created region files of all identified point sources and their associated fluxes to mask them from the analysis. The exact procedure is described in more detail in section \ref{sec:mod:agn}.

Finally, we reprojected all the event data and exposure maps to a common tangent point and co-added them. We use these combined files to measure the number of counts, surface brightness, and particle background levels across the field in the rest of the analysis.

\section{Model and Method}
 \label{sec:model}
Our forward-model includes projected ICM emission, two types of astrophysical backgrounds (the Galactic foreground and unresolved AGN), as well as the particle background. The forward-model approach has several advantages, namely that 1) additional source or background components can easily be added to the model, 2) by modeling all background components we can robustly account for their distributions, rather than just subtracting a single value or assuming Gaussianity, and 3) by bringing the model to the data, the Poisson noise is handled appropriately, even in bins with low number of counts. This forward-model approach thus constitutes a novel and statistically rigorous way to measure ICM fluctuations in galaxy clusters. In the following subsections, we give an overview of the spatial layout, the way we derive the turbulent velocity and clumping facDtor from surface brightness fluctuations, and the individual model components and statistical implementation.

\subsection{Spatial layout}

\begin{figure*}
\centering
\includegraphics[width=0.42\linewidth]{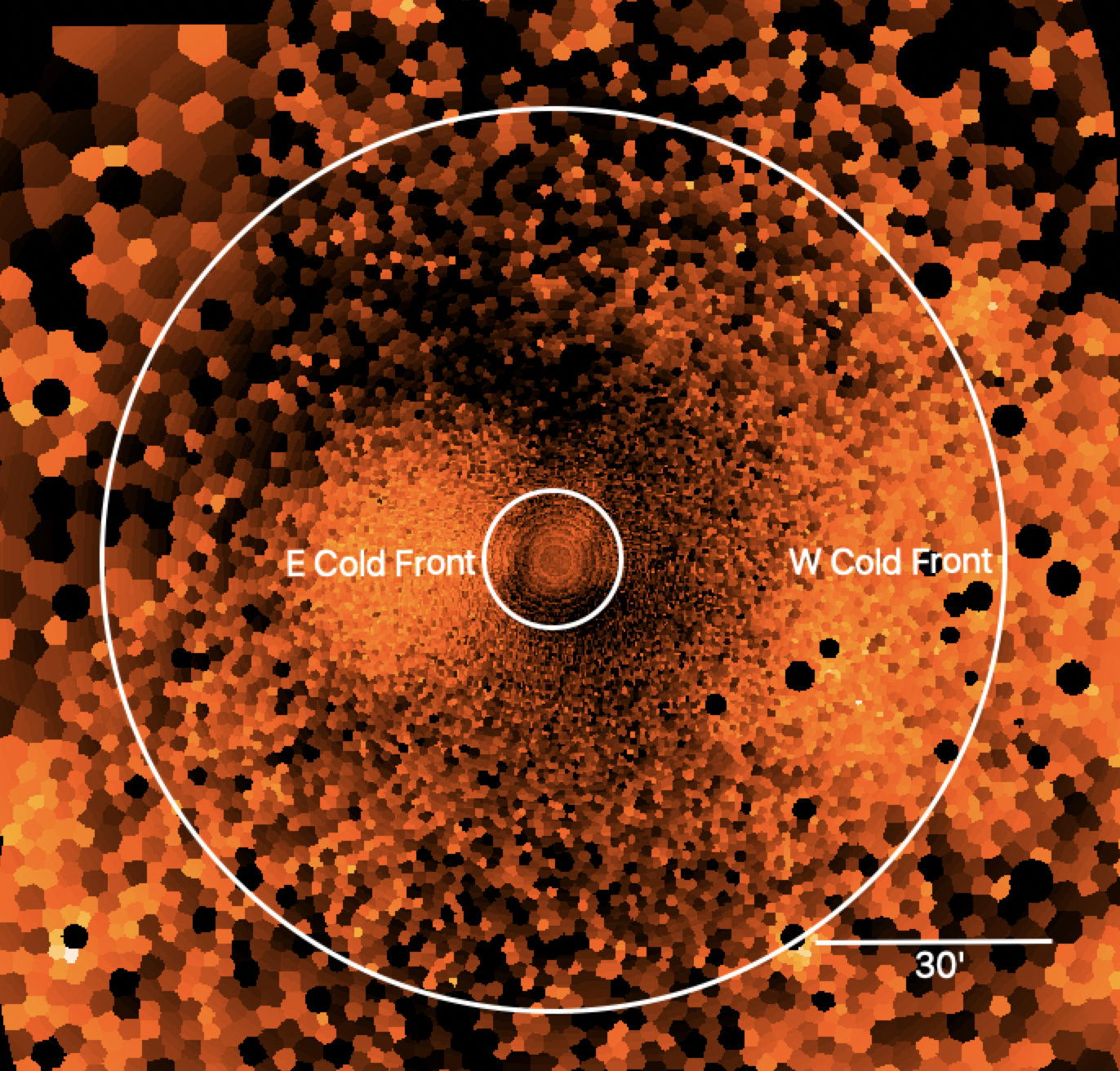} 
\includegraphics[width=0.42\linewidth]{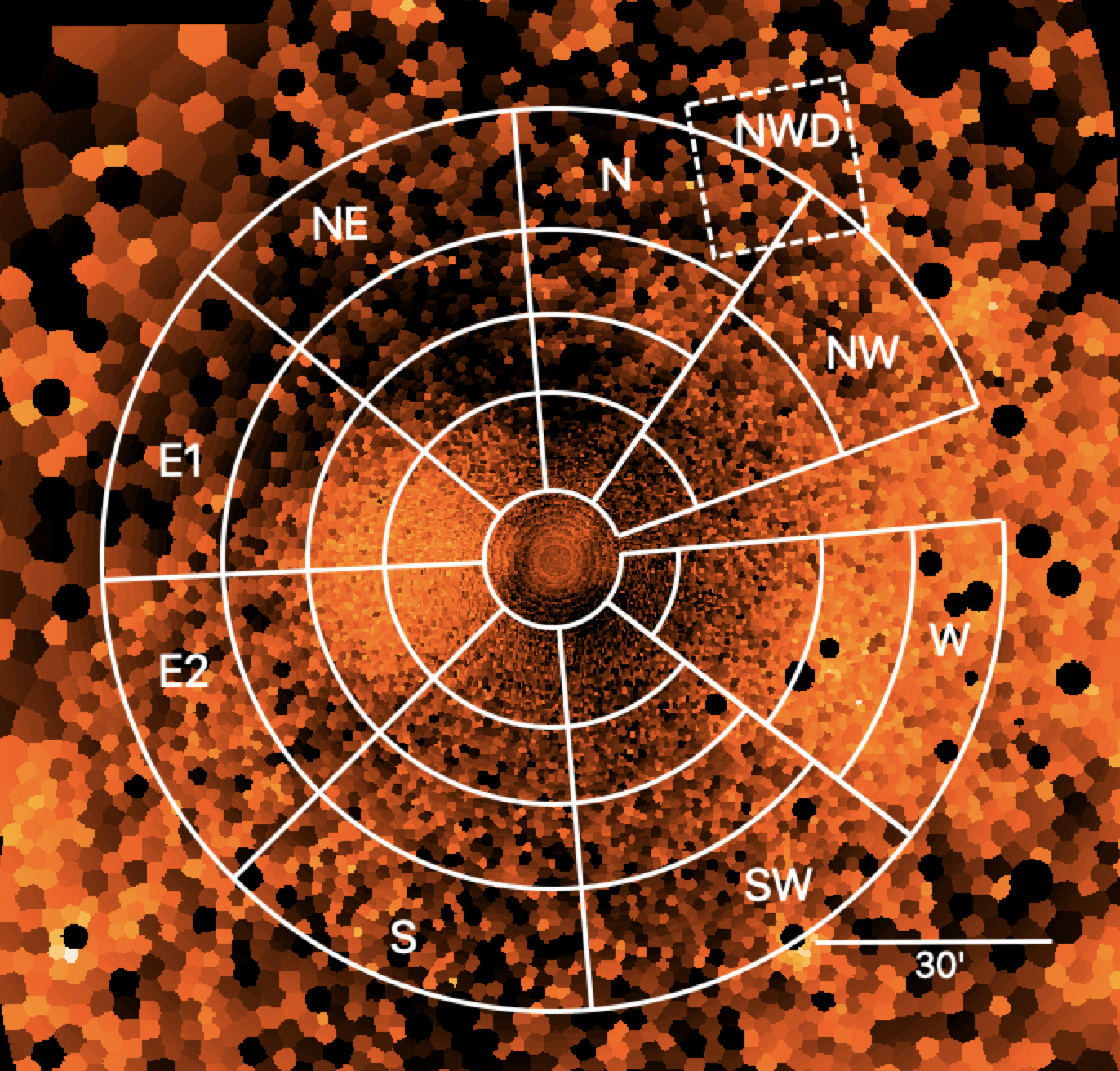} 
\includegraphics[width=0.845\linewidth]{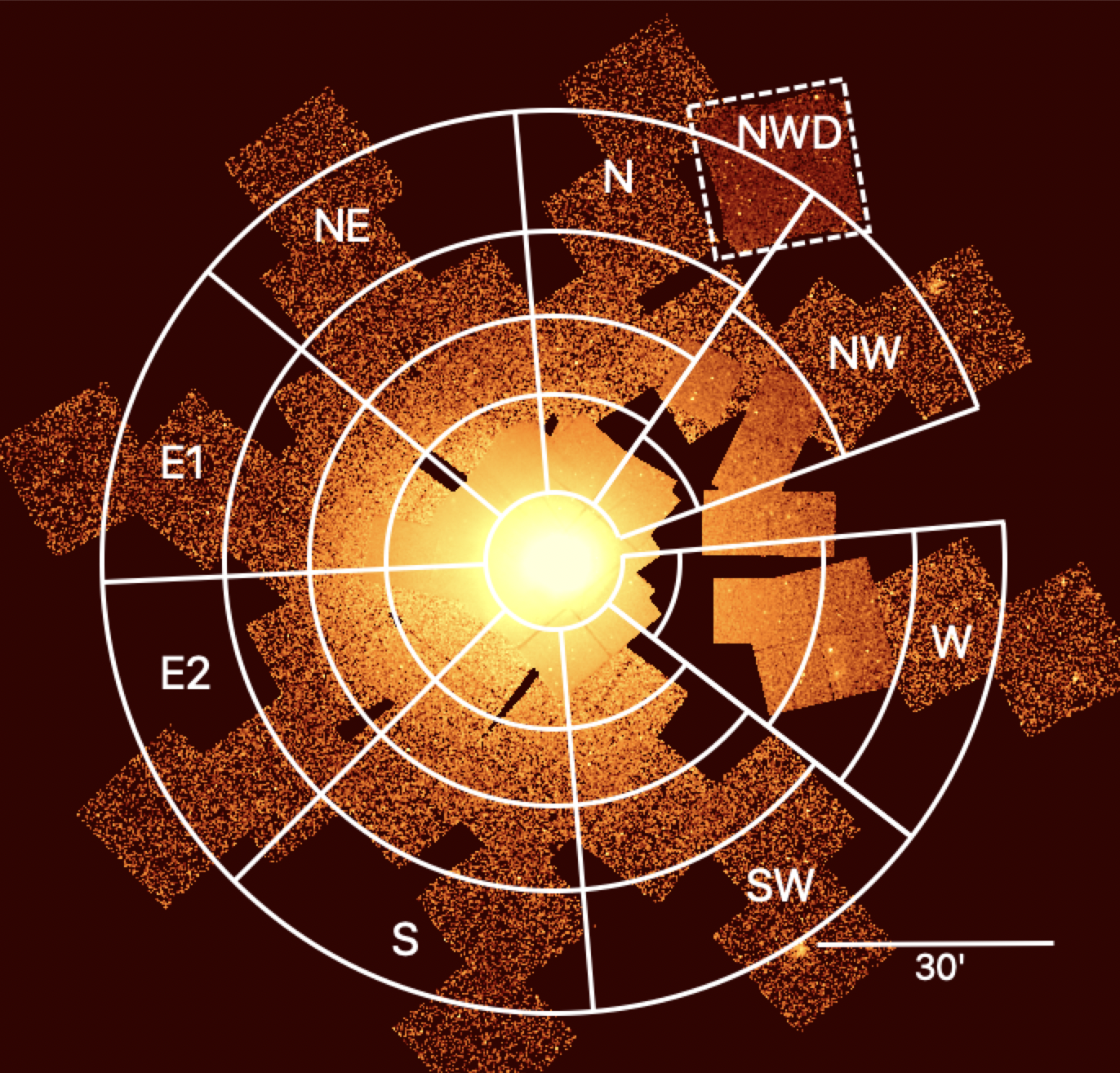} 
\caption{The spatial layout of the shells in the Perseus field. {\it Top left panel:} A ROSAT residual surface brightness image with a symmetrical beta model subtracted \citep{Simionescu2012}, showing the cold fronts created by the sloshing motions of the cluster. The annulus indicates the inner and the outer radii considered in our analysis, $0.3\,r_{2500}$ and $2.2\,r_{2500}$ {\it Top right panel:} The ROSAT residual image with the sectors overlaid, showing the 28 shells considered in the analysis, {\it Bottom panel:} Exposure-corrected {\it Chandra} mosaic of the used Perseus observations, with the sectors overlaid.}

\label{fig:arms}
\end{figure*}

\begin{figure}
\centering
\includegraphics[width=0.99\linewidth]{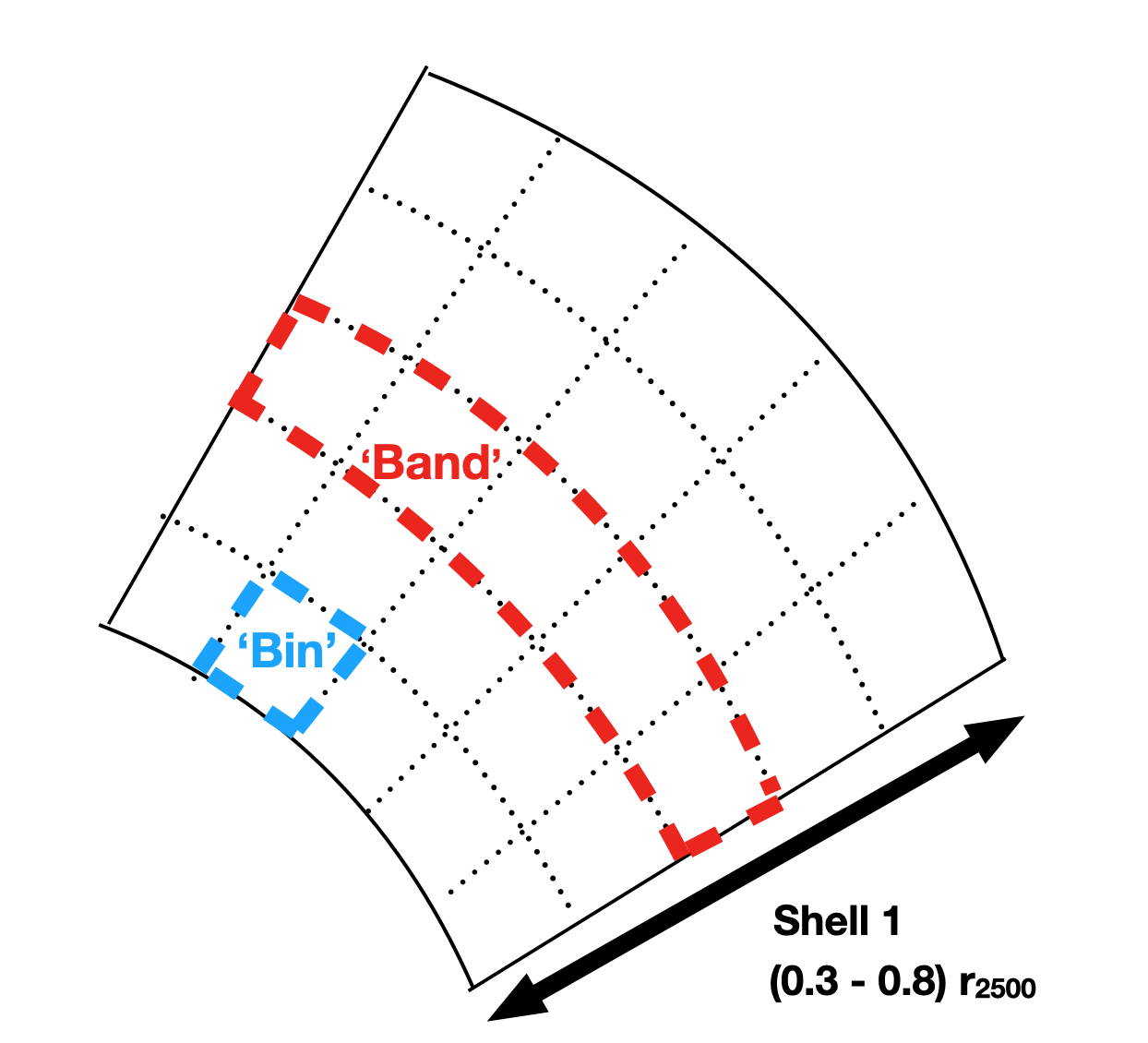} \\
\caption{Schematic overview of the region layout within a shell. Each shell is subdivided into approximately square-sized bins with a surface area of 1 arcmin$^{2}$. The bins are laid out along annuli, using the Perseus cool core as the center. To account for the surface brightness gradient, the cluster emission component of each set of bins at the same radius (a band) is modeled as a separate log-normal distribution, while the log-normal standard deviation $\sigma$ is shared between all bands in the shell. Note that in this schematic layout all bands contain 6 bins, but in practice the number of bins will vary based on how many 1 arcmin$^{2}$ fit in the band at a given radius. The black arrow indicates the radial extent of the shell, in this example $(0.3-0.8) r_{2500}$. }
\label{fig:schem}
\end{figure}

\begin{table}
\centering
\caption{Overview of the sectors and the number of shells within each sector. The shells in the E1 and E2 sectors at the same radius are analyzed together as a single shell. }
\label{tab:armshells}
\begin{tabular}{c  c  c  c}
\hline \hline
Sector & Azimuth E of N & shells & Shell radii [$r_{2500}$] \\ \hline
NE & $5\degree$-$50\degree$ & 4 & {\bf 1)} 0.3--0.8, {\bf 2)} 0.8--1.2 \\
& & & {\bf 3)} 1.2-1.6, {\bf 4)} 1.6--2.2 \\
E1 & $50\degree$-$92.5\degree$ & 4 & as NE\\
E2 & $92.5\degree$-$135\degree$ & 4 & as NE \\
S &$135\degree$-$185\degree$ & 4 & as NE\\
SW &$185\degree$-$232.5\degree$ & 4  & as NE\\
W & $232.5\degree$-$275\degree$ & 4 & {\bf 1)} 0.3--0.6, {\bf 2)} 0.8--1.3 \\
& & & {\bf 3)} 1.3--1.8, {\bf 4)} 1.8--2.2 \\
NW & $290\degree$-$325\degree$ & 3 & {\bf 1)} 0.3--0.6, {\bf 2)} 0.7--1.5\\
 & & & {\bf 3)} 1.5--2.2 \\
NWD & $317\degree$-$341\degree$ & 1 & {\bf 1)} 1.7--2.7 \\
N &  $325\degree$-$365\degree$ & 4 & as NE \\  \hline
\end{tabular} \\
\end{table}

We aim to measure the surface brightness fluctuations in self-contained regions across the Perseus field. In order to do so, we have divided the Perseus field into several sectors, following three criteria: 1) the azimuthal opening angle is no larger than $50 \degree$, in order to minimize the effect of non-radial surface brightness gradients caused by the overall approximately elliptical surface brightness distribution and apparent large-scale sloshing motions \citep{Simionescu2012}, 2) there is an as-clean-as-possible division between `quiescent' sectors along the north-south axis, and `non-quiescent' sectors positioned on the most prominent cold fronts on the east-west axis, and 3) the sectors follow the layout of \textit{Chandra} CCD observations, such that the regions in each sector are as contiguous as possible. As the cluster center, we use the coordinates of \cite{Simionescu2012}: $\alpha$=3:19:47.7, $\delta$=+41:30:41.9.

The \textit{ROSAT} residual surface brightness map of \cite{Simionescu2012} was used to identify the location of the cold fronts within Perseus. An overview of the cold fronts in the ROSAT residual map, the \textit{Chandra} ACIS observations, and the sectors is shown in Figure \ref{fig:arms}. Along the cluster's E-W major axis, where the most obvious cold fronts are present, we define 3 sectors, subdividing the eastern direction into 2 (E1 and E2) in order to minimize the azimuthal opening angle. We also define 5 `quiescent' sectors: S, SW, NE, N, NW, which are oriented away from the cold fronts (though which may, in detail, still exhibit varying degrees of quiescence). In each of these 7 sectors, we restrict ourselves to minimum and maximum radii of $0.3-2.2$\,$r_{2500}$. Additionally, we analyze a separate `deep-field' set of 4 observations (ObsIDs 13989 to 13992) in the northwestern region with a total exposure time of $148\,$ks, which we refer to as the NWD region. 

Each of these sectors is then subdivided into a number of "shells", each spanning a particular radial range, within which we will infer the local turbulent velocity and clumping factor. Each shell is further divided into a rough grid of $1'\times1'$ approximately square "bins", which are the regions used to determine the strength of the surface brightness fluctuations within a shell. Crucially, the bins are defined without reference to the surface brightness, and the number of counts within them therefore follows Poisson statistics. This is not the case for adaptive binning algorithms such as Weighted Voronoi Tessellations or contour binning. Because all the model components are forward-modeled towards the observed data, our model is able to deal with Poisson noise even when a bin contains zero counts, and thus there is no need to ensure that each bin contains a minimum number of counts. We note that the results are only lightly sensitive to the chosen bin size of $1'\times1'$. As we will see in section \ref{sec:disc:3D}, larger scales (closer to the size of the shell itself) dominate the result in any given shell. 

A schematic overview of the layout within a shell is shown in Figure \ref{fig:schem}. Within each shell, the bins are laid out along annuli using the same cluster center for all shells. The shells themselves are chosen to have $>100$ bins, such that meaningful inferences can be made about the variance of the cluster emission. For consistency between the sectors, most use the same shell radii. The exceptions are the W and NW sectors, where gaps in the coverage combined with a comparatively small azimuthal ranges required small adjustments. An overview of the sectors and the number of shells within each sector is given in Table \ref{tab:armshells}.

The eastern sector represents a special case: as the eastern cold front extends over a relatively large angle, we have subdivided it into the E1 and E2 sectors. In practice, this means that a band of bins in the E1 and E2 sectors at the same radius are allowed to have different mean surface brightness. In doing so, we can correct for the fact that the global surface brightness profile of Perseus is not spherical over a large angle. As such, the E1 and E2 sectors are binned separately but subsequently analyzed together, and the results in Section \ref{sec:results} are listed for the E sector as a whole. 

The size of our bins is  $1'\times1'$, while the width of the \textit{Chandra} PSF at energies $<3.5$ keV remains below $\sim 5$ arcsec across the majority of the field. Because the PSF is significantly smaller than the bin size, we have chosen to ignore PSF effects throughout the analysis. In particular, the point sources that are incorporated in the modeling (see Section \ref{sec:mod:agn}) are always assumed to lie within a single bin.

\subsection{Cluster emission, clumping and turbulence}
\label{sec:mod:eqs}

For turbulent motions in an inviscid, isothermal gas, the equations governing the gas motions can be described only in terms of the log of the density $\log{\rho}$. Under the assumption that the perturbations in the gas are random and uncorrelated, it can readily be seen that the density PDF should be log-normal in shape. Although this is a simplification, both analytical and numerical models of turbulent motions in such gases show that a log-normal distribution is a good approximation of the density PDF \citep{Passot1998, Nordlund1999}.

Log-normality appears to be a valid approximation in real observations of the ICM as well. It has been shown observationally that, after correcting for the radial surface brightness gradient, the surface brightness distribution in a given region of a galaxy cluster is approximately log-normally distributed \citep[][]{Kawahara2007, Kawahara2008, Khedekar2013}. In simulations of galaxy clusters, \cite{Zhuravleva2013} find that the density PDF's can be separated in a log-normally distributed bulk component and a high density tail. This high-density tail is at least partly a simulation artifact, which arises because cold and dense gas in subhalos is more prominent in cooling and star formation (CSF) simulations \cite{Zhuravleva2013}. 

The radii of interest in this paper, $(0.3-2.2)\,r_{2500}$, are outside the cool core where the AGN feedback and other astrophysical processes play an important role in the gas dynamics, but well within the virial radius. We thus use the log-normal model to describe the cluster density PDF. These log-normal fluctuations are on top of the radial gradient of the cluster, which we account for in the model. As we will see in Section \ref{sec:results}, the log-normal model for the cluster emission provides a good description of the data. 

We can link fluctuations in the density distribution to fluctuations in the emissivity distribution through equation 4 of  \cite{Zhuravleva2016}:
\begin{equation}
    \left(\frac{\delta \epsilon}{\epsilon}\right)_i = \left(\frac{\delta 
    \rho}{\rho}\right)_i \left[2 + (\zeta_i -1) \frac{d \ln \Lambda(T)}{d \ln T} \right],
\end{equation}
where $\delta \epsilon/\epsilon$ denote the emissivity fluctuations, $\delta \rho/\rho$ denote the density fluctuations, the subscript $i$ refers to the type of fluctuation (isobaric, adiabatic, or isothermal), and $\zeta_i=0, 5/3, 1$ for isobaric, adiabatic, and isothermal fluctuations respectively.  In the $0.6$ to $3.5$ keV  energy range and for gas temperatures $>3$ keV, we can simplify this by assuming that $\frac{d \ln \Lambda(T)}{d \ln T} \sim 0$ \citep{Zhuravleva2016}:
\begin{equation}
    \left(\frac{\delta \epsilon}{\epsilon}\right) \approx 2 \left(\frac{\delta \rho}{\rho}\right).
    \label{eq:sbn}
\end{equation}
For this reason, as well as to make sure the \textit{Chandra} PSF remains small enough to not affect the analysis, we restrict ourselves to the 0.6--3.5 keV energy band.

If we assume that the cluster emission is log-normally distributed, the definition for the clumping factor (equation \ref{eq:clumpf}) can be re-written only in terms of the density log-normal standard deviation $\sigma_{\rho}$ (see Appendix A of \cite{Eckert2015} for the derivation):
\begin{equation}
    C = \exp{({\sigma_{\rho}}^{2})}.
    \label{eq:clumpsig}
\end{equation}

Additionally, for relaxed clusters, the fluctuations in the density distribution at a given length scale, $k$, can be linked to the one-component turbulent velocity, $V_{k, 1D}$, at that length scale:
\begin{equation}
    \left(\frac{\delta \rho_k}{\rho}\right) = \eta_1 \frac{V_{k, 1D}}{c_s}.
    \label{eq:vturb}
\end{equation}

The factor $\eta_1$ is a proportionality constant $\approx 1 \pm 0.3$ \citep{Zhuravleva2014}. We assume that this relationship holds for Perseus as well, given that it is a relaxed cluster according to the Symmetry-Peakiness-Alignment (SPA) criterion of \cite{Mantz2015}.

We note that equation \ref{eq:vturb} is calibrated using non-radiative cosmological simulations which do not include physical processes such as cooling, star formation, and AGN and supernova feedback. We assume that at the radii investigated in these paper, the effects of these processes are small. The cooling time of the ICM at $r>0.3r_{2500}$ is of order 10 Gyr or greater \citep{Dunn2006}, and the effects of star formation and SN feedback are expected to be most important closer to center, where the stellar density is highest. \cite{Zhuravleva2022} have studied the proportionality between density and velocity perturbations in hydrodynamical simulations including additional physical effects (cooling, star formation, AGN and SN feedback, and the UV background). They find that while including these effects increases the scatter of the relationship, the mean relationship is unchanged within the uncertainties.

By using equations \ref{eq:sbn}, \ref{eq:clumpsig}, and \ref{eq:vturb}, we can thus directly connect fluctuations in the emissivity distribution to two quantities: the clumping factor and, assuming it is turbulent motions that sources these fluctuations, the one-component turbulent velocity. 

Finally, we require a way to link the 3-dimensional quantity of emissivity to the 2-dimensional quantity that is observed in the data: the surface brightness. In general, projection has the effect of suppressing the amplitude of observed fluctuations, but the exact level of suppression depends on the surface brightness and density profiles of the cluster \cite{Churazov2012}. In order to measure this suppression factor, we perform a power spectrum analysis on a select few regions, which we will discuss further in Section \ref{sec:disc:3D}.

\subsection{Model components}
\label{sec:mod:comps}
The model consists of four additive components. In order of importance these are: the projected cluster emission, the particle background, the background from unresolved AGN, and the Galactic foreground. For each component, we find the probability distribution of the corresponding expected number of counts $\Lambda$, as described in the subsections below. For all components except the particle background, we use the \textit{Chandra} exposure maps to convert between surface brightness and number of counts. The expected number of counts for each individual component are then combined in the likelihood equation in Section \ref{sec:mod:ll}. Throughout this section, we use the symbol $\sim$ to mean 'follows the distribution of'.

\subsubsection{Cluster emission}
\label{sec:mod:cls}
As detailed above, we assume that the cluster emission in each 1 arcmin$^2$ bin follows a log-normal surface brightness distribution. This log-normal distribution has the free parameters $\mu$ and $\sigma_f$, where the standard deviation $\sigma_f$ is shared between all the bins in the shell, and the mean $\mu$ is shared between all the bins in the same band. The log-normal distribution is described in terms of photon surface brightness, and $\mu$ thus has the units of photon\,cm$^{-2}$\,s$^{-1}$\,arcsec$^{-2}$. 

For a given bin, the mean number of cluster emission counts $\Lambda_C$ is distributed as
\begin{equation}
\label{eq:clslognorm}
    {\Lambda}_{\rm C} \sim \rm LogNorm(\mu_{\rm band} + 
    \ln{E_{\rm bin}}, \sigma_{f, {\rm shell}}),
\end{equation}
 where $\mu_{\rm band}$ is the mean surface brightness in that band of spatial bins, $\sigma_{f, {\rm shell}}$ log-normal standard deviation of the surface brightness distribution in the shell, and $E_{\rm bin}$ the total value of the exposure map in that bin, with units of ${\rm cm}^{2} \, {\rm counts}^{-1} \, {\rm photon}$. 

\subsubsection{Particle background}
\label{sec:mod:pb}

Using the ACIS 'stowed' background observations, we can calculate the expected number of particle background counts in a given region of the science observation.

Let $P$ be a random variable for the number of counts in the stowed background observation:
\begin{equation}
P \sim \rm{Poisson}(\Lambda_P),
\end{equation}
where $\Lambda_P$ is expected number of counts in the background observation. The number of expected counts in the science observation in an equivalent region is then $\Lambda_B = \Lambda_P \tau$, where $\tau$ is the ratio of exposure times of the science observation and the background observation $\tau=t_S/t_B$. 
If we choose a Gamma distribution as a prior with shape $k_B$ and scale $\theta_B$, then the expected number of counts $\lambda_P$ in the background observation are distributed as:
\begin{equation}
\label{eq:bcounts}
\Lambda_P \sim \rm{Gamma}\left(k_B + \hat{B}, 
\frac{\theta_B}{1 + \theta_B} \right).
\end{equation}
We can then obtain the predicted number of particle background counts in the science observation by multiplying $\Lambda_P$ with the ratio of exposure times $\tau$. We choose a uniform prior ($k_B = 0.5, \theta_B \to \infty$). Because our method directly uses the number of counts from the stowed background, no vignetting correction is necessary, just as in the more common case of the 'blank-sky' background.

Each shell in our analysis is a mosaic of different observations with different exposure time, and each observation might also have a different level of particle background. We apply this particle background scaling on a bin-by-bin basis by modulating the exposure time ratio $\tau$ into an effective ratio $\tau_{\rm eff}$. For each chip of each ACIS observation, we calculate an energy scaling factor by comparing the exposure time-scaled counts in the data and the stowed background mosaic in the $9.5-12.0$ keV range. At these energies, the effective area of \textit{Chandra} is small enough that all detected counts are expected to be of particle background origin. For each individual ObsID, we calculate a background scaling factor $f_{\rm bgscal}$ and an error on the scaling $\sigma_{\rm bgscal}$. Given that the number of $9.5-12$ keV counts on a single chip are $>200$ in all science and background observations, we can safely approximate the errors on the scaling factor to be Gaussian.

We then apply the energy scaling by modulating the exposure ratio $\tau$. For each spatial bin within the shell, we determine which observations overlap with that bin, and then calculate the effective exposure ratio for that bin as
\begin{equation}
\label{eq:taueff}
    \tau_{\rm eff} = (t_{\rm S1} + t_{\rm S2} + ...)  \left(\frac{t_{\rm B1}}{F_{\rm bgscal_1}} + \frac{t_{\rm B2}}{F_{\rm bgscal_2}} + ....\right)^{-1},
\end{equation}
where the indices 1,2.. represent each observation that bin overlaps with. In the above equation, the background scaling factor is itself a random variable that depends on the measured background scaling and its associated error:
\begin{equation}
\label{eq:bgdraw}
    F_{\rm bgscal} \sim {\rm Normal}(f_{\rm bgscal}, \sigma_{\rm bgscal}).
\end{equation}
The error on the background scaling factor typically is around $5\%$ for the short $5\,$ks observations. In order to minimize computing time, we opted to ignore the error on the background scaling in the modeling and keep it as a fixed parameter, having verified in the four shells of the N sector that this does not significantly affect the results.

The effective exposure ratio from equation \ref{eq:taueff} can be used together with equation \ref{eq:bcounts} to compute the expected number of particle background counts $\Lambda_B$ for each bin. 

\subsubsection{AGN background}
\label{sec:mod:agn}

Both resolved and unresolved AGN can contribute to the total number of observed counts in a given bin. Resolved AGN can be masked from the data and excluded from the rest of the analysis, but the contribution of unresolved and therefore undetected AGN must be modeled. Within the data, we identify the resolved point sources in the field using the Cluster AGN Topography Survey pipeline (CATS; Canning et al. in prep). For each of the point sources identified with CATS, we estimate the flux from the number of counts assuming a power law spectrum with photon index $\Gamma=1.5$.

As part of the CATS pipeline, sensitivity maps for each CCD of each observation are also calculated, which indicate the minimum flux at which an AGN can be detected at each position. We use the sensitivity maps from the $\approx 5$\,ks observations to identify a flux limit above which $>95\%$ of AGN are detected, even in these relatively short exposures and at off-axis pointings. The CATS sensitivity maps are thus used to determine at which flux we treat AGN as 'resolved' or 'unresolved' point sources. We find that for the $\approx 5$\,ks observations in the Perseus data set, the $95\%$ sensitivity level does not exceed $2 \times10^{-6}$ photon\,cm$^{-2}$\,s$^{-1}$. We therefore choose an AGN flux cutoff of $4 \times10^{-6}$ photon\,cm$^{-2}$\,s$^{-1}$, a factor two above this sensitivity limit, in order to ensure that no AGN above the flux cutoff remain undetected by CATS. All AGN above the cutoff are masked \footnote{Given that AGN with low flux are accounted for in the model rather than masked from the analysis, we note that our results are not sensitive to the details of the CATS pipeline. Comparable results might be achieved by straightforward application of standard source detection tools such as {\it wavdetect}.}.

\begin{figure}
\vskip -0.5cm
\centering
\includegraphics[width=0.99\linewidth]{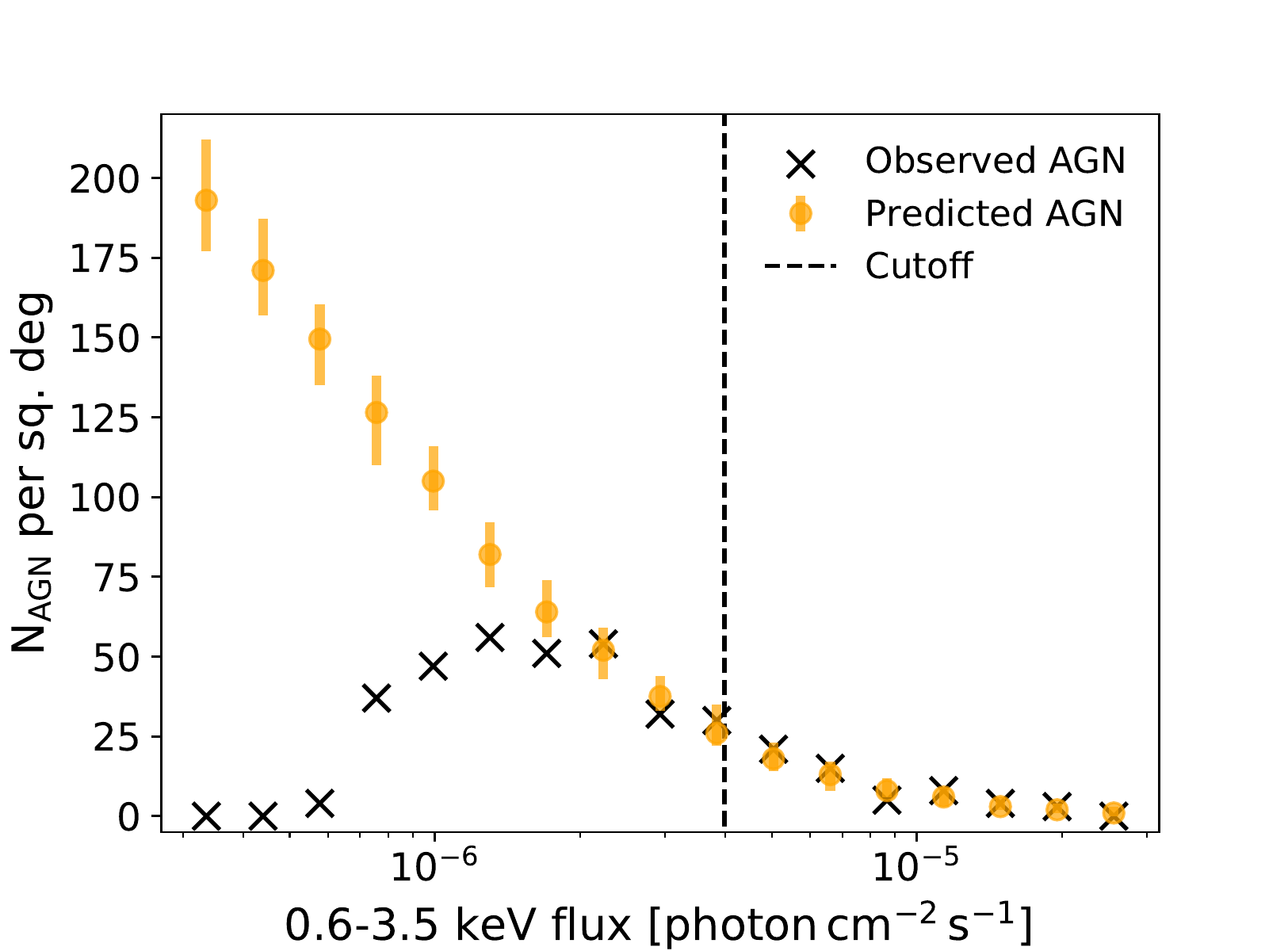} \\
\caption{Observed versus predicted number of AGN as a function of 0.6--3.5\,keV flux over an area of 1 square degree. The dashed line indicates the cutoff above which we excise the AGN from the analysis. Below the cutoff, the AGN are modeled using the redshift-integrated AGN flux distribution predicted from the AGN luminosity function.}
\label{fig:agnpred}
\end{figure}

In order to model the unresolved AGN, we use models of the AGN luminosity function $\Phi(L, z)$, which describes the AGN number density at a luminosity $L$ and redshift $z$. Using this function and the model parameters of \cite{Miyaji2015}, we calculated the predicted number density of AGN integrated over all redshifts and within a given flux interval through a Monte Carlo process. We compared the distribution of thusly predicted AGN with the number of detected AGN in Perseus with the CATS pipeline. Figure \ref{fig:agnpred} shows the observed AGN number density as a function of $0.6-3.5$ keV photon flux (photon\,cm$^{-2}$\,s$^{-1}$), versus the model-predicted AGN number density over a solid angle of 1 square degree. At fluxes $>2 \times10^{-6}$ photon\,cm$^{-2}$\,s$^{-1}$, the number of predicted AGN matches very well to the number of detected AGN. This shows that the AGN luminosity function provides a good description of the AGN number density in the Perseus field, and that our choice of a cutoff at $4 \times10^{-6}$ photon\,cm$^{-2}$\,s$^{-1}$ is appropriate.

As with the point sources found by the CATS pipeline, we assume that the typical AGN spectrum is a power law with photon index $\Gamma=1.5$, so that we can estimate the total flux from AGN in a given flux range from the number density. The expected number of counts from unresolved AGN in a bin $\Lambda_{\rm A}$ is then given by integrating the redshift-integrated AGN luminosity function over a flux interval. From this function, we can calculate the expected number of AGN counts in a given region:

\begin{equation}
\label{eq:AGNfunc}
{\Lambda}_{\rm A} \sim \phi (F_{\rm min}, F_{\rm max}, \alpha_{\rm bin}) E_{\rm bin}
\end{equation}
where $\phi$ is the PDF of the flux (in units of photon\,cm${-2}$\,s$^{-1}$) from all unresolved AGN within the flux range $F_{\rm min}$ to $F_{\rm max}$, and in the surface area $\alpha_{\rm bin}$. In our implementation, we use $F_{\rm min} = 0$, and $F_{\rm max} = 4 \times10^{-6}$ photon\,cm$^{-2}$\, s$^{-1}$, corresponding to the cutoff in Figure \ref{fig:agnpred}. 

\subsubsection{Galactic foreground}

\begin{table}
\centering
\caption{Model parameters for the APEC + PHABS*APEC model fit to the Galactic foreground.}
\label{tab:galf}
\begin{tabular}{c  c  c}
\hline \hline
Parameter & APEC & PHABS*APEC2 \\ \hline
kT [keV] & 0.0974 & 0.221 \\
Z/Z$_\odot$ & 1 & 1 \\
Norm$^1$ & $4.80 \times 10^{-7}$ & $ 2.32 \times 10^{-6}$ \\
$n_H$ [cm$^{-2}$] &  & $1.36 \times 10^{21}$ \\ \hline
\end{tabular} \\
$^1$ APEC normalization in the default \textsc{XSPEC} units of $\frac{10^{-14}}{4\pi [D_A (1 +z)]^2} \int n_e n_H dV$. 

\end{table}

In order to model the Galactic foreground, we use Perseus observations from the ROSAT All-Sky Survey. We defined 8 circles with radius of $1 \degree$, $2 \degree$ away from the center of Perseus. The regions to the W and NW were contaminated by nearby sources, while the remaining six regions were used to model the Galactic foreground. 

As a spectral model, we fit a combination of an absorbed and unabsorbed thermal component: \textsc{APEC + PHABS * APEC} in \textsc{XSPEC}. These two model components reflect thermal emission from the 'Local Bubble', and emission from Galactic halo respectively \citep[for a discussion on \textit{Chandra's} soft X-ray background, see][]{Hickox2005}. An absorbed power law component was included in the fit to model the contribution from unresolved AGN, but this component is modeled separately in our analysis. The abundance Z and the galactic absorption $N_H$ were fixed, following \cite{Mantz2022}. A fit with a third thermal component of $\sim 0.6 keV$ was also attempted \citep[see e.g.]{Urban2014, Bluem2022}, but this did not result in an improved fit. The results of the fit are shown in Table \ref{tab:galf}.

We use the parameters in Table \ref{tab:galf} to calculate the $0.6-3.5$ keV surface brightness of the Galactic foreground: $2.13 \times 10^{-10}$ photons cm$^{-2}$ s$^{-1}$ arcsec$^{-1}$. This is taken to be a constant in the model.  The number of predicted counts from the Galactic foreground in a given bin is then

\begin{equation}
    \Lambda_{\rm g} = 2.13 \times 10^{-10} E_{\rm bin}.
\end{equation}

\subsection{Likelihood function}
\label{sec:mod:ll} 

For a given shell, we aim to constrain the variance of surface brightness fluctuations, $\sigma_{\rm shell}$, and the mean brightness for each band ($i$) within the shell, $\mu_i$, using the number of 0.6--3.5 keV counts measured in each bin ($j$) in the shell, $k_j$.
The likelihood for each bin is Poisson, conditional on the corresponding predicted expectation value of the number of counts, which we model as the sum of the 4 components discussed above: $\Lambda_j = \Lambda_{\rm C}(\mu_{i(j)}, \sigma_{\rm shell}) + \Lambda_{{\rm B},j} + \Lambda_{{\rm A},j} + \Lambda_{{\rm g},j}$, where $i(j)$ indicates the band containing the $j$th bin.
In practice, we directly marginalize over the latent parameters $\Lambda_\mathrm{C}$, $\Lambda_\mathrm{B}$, $\Lambda_\mathrm{A}$ and $\Lambda_\mathrm{g}$ using Monte Carlo integration, such that the likelihood for a bin is
\begin{equation}
    \mathcal{L}_j = \left\langle \frac{\lambda_j^{k_j} e^{-\lambda_j}}{k_j!} \right\rangle,
\end{equation}
where $\lambda_j$ represents a random realization of $\Lambda_j$ based on the PDFs described in the previous section, and the angled brackets represent an average over such realizations (we find that 10,000 provides a good compromise between speed and precision).
The complete likelihood for the shell is simply the product of $\mathcal{L}_j$ over bins within the shell.
In order to sample the posterior distributions of the remaining parameters, $\sigma_{\rm shell}$ and ${\mu_i}$, we use the Markov Chain Monte Carlo sampler of \cite{Goodman2010}, implemented through the Python package {\sc emcee}. 

\section{Results}
\label{sec:results}
\begin{figure*}
\centering
\includegraphics[width=0.99\linewidth]{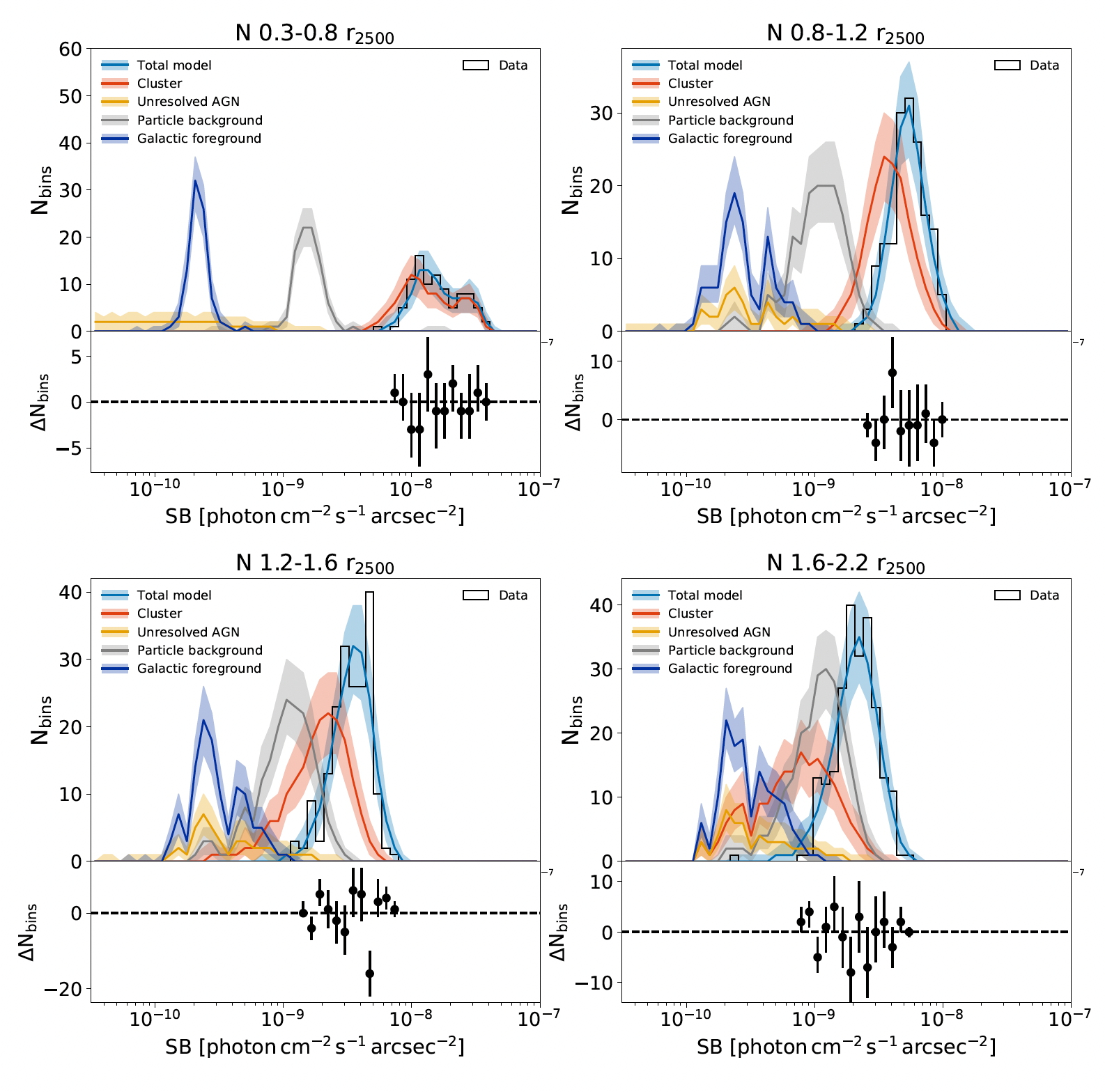} \\
\caption{Forward-simulated model histograms resulting from the MCMC-sampled posterior distributions, compared with the observed 0.6--3.5\,keV surface brightness data in the 4 shells of the northern sector. The 'total model' represents the sum of the 4 individual model components: cluster emission, unresolved AGN, particle background and Galactic foreground. The model error envelopes were created by repeating the forward simulation process 10,000 times, and taking the 14th and 86th percentiles at each histogram bin as the lower and upper model boundaries. We note that the residual plots below each panel are purely meant as a visual aid, as the errors are correlated. }
\label{fig:modelvis}
\end{figure*}

\begin{figure*}
\centering
\vspace*{-4mm}

\includegraphics[width=0.95\linewidth]{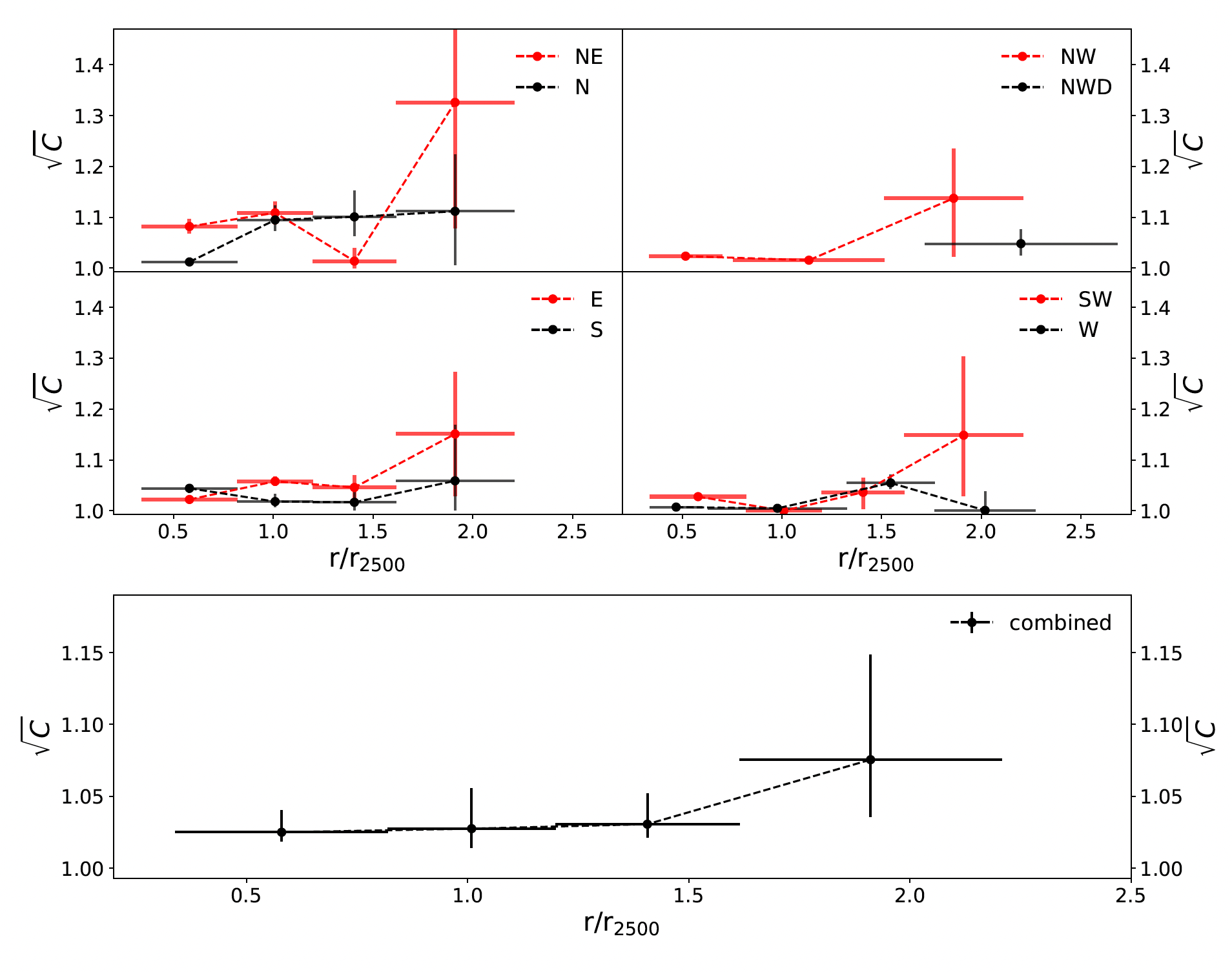} \\
\vspace*{-4mm}
\caption{The density bias $\sqrt{C} - 1$ measured with 1 x 1 arcmin$^2$ regions, in each of the 8 sectors in Perseus. Plotted are the statistical errors, which indicate the $68.3$ per cent highest posterior density interval (HDPI), with the mode listed as the central value. The combined results were obtained by fitting a Gaussian mean plus a scatter to the results of individual shells: see text for details. }

\label{fig:clump}
\end{figure*}

\begin{figure*}
\centering
\vspace*{-4mm}

\includegraphics[width=0.95\linewidth]{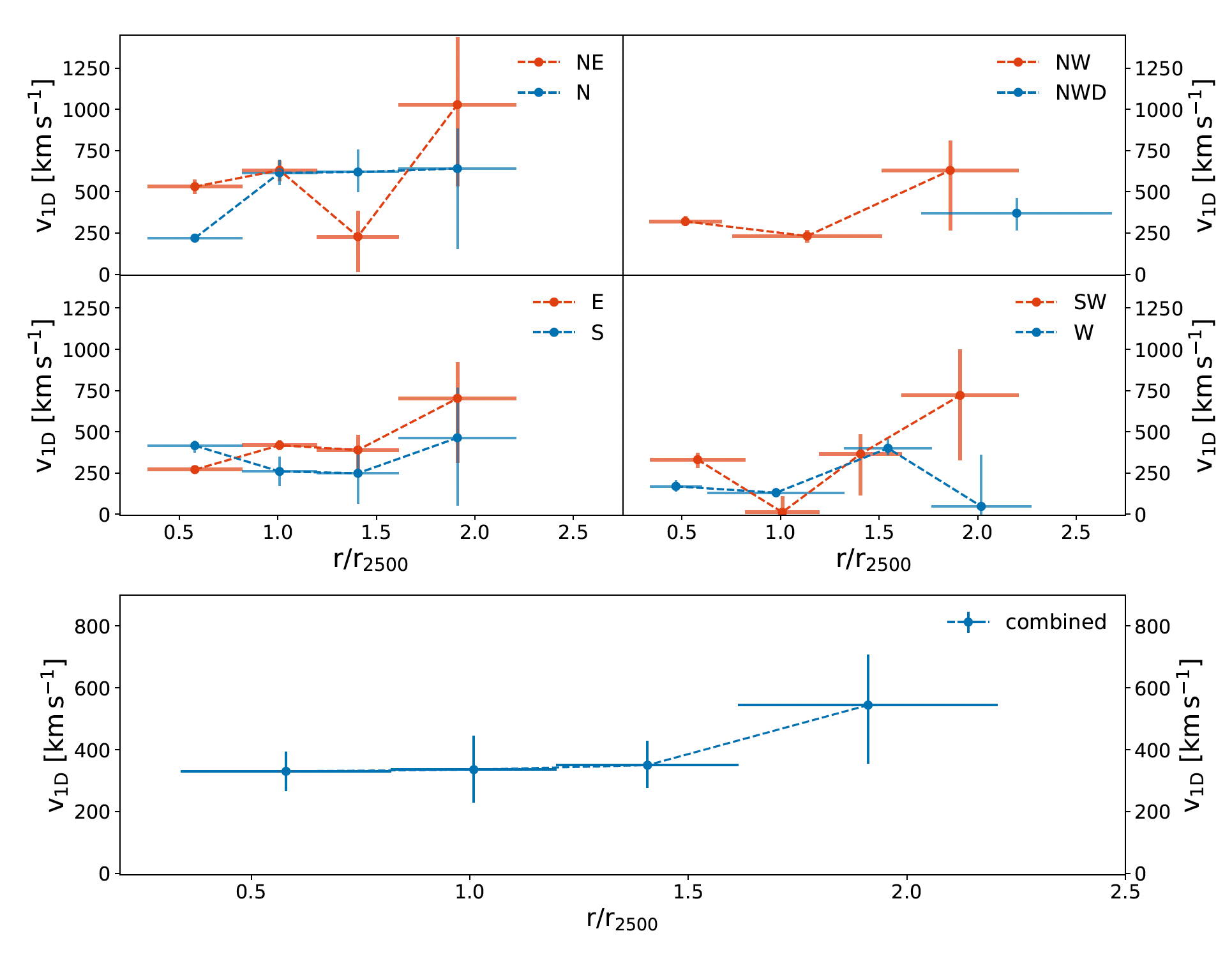} \\
\vspace*{-4mm}
\caption{The one-component turbulent velocity measured with 1 x 1 arcmin$^2$ regions, in each of the 8 sectors in Perseus. Plotted are the statistical errors, which indicate the $68.3$ per cent highest posterior density interval (HDPI), with the mode listed as the central value. The combined results were obtained by fitting a Gaussian mean plus a scatter to the results of individual shells: see text for details. }

\label{fig:vturb}
\end{figure*}

{\renewcommand{\arraystretch}{1.35}

\begin{table}
\centering
\caption{Full results of the MCMC analysis. Listed are the $68.3$ per cent highest posterior density intervals (HPDI) of $\sigma_{f}$, with the mode of each distribution as the central value, the plasma temperature used for computing the soundspeed (see Appendix \ref{apx:temp}), and the derived quantities $v_{k, 1D}$ and density bias $(\sqrt{C}$. The 'Comb' sector represents the fitted average for the NE, E, S, SW, W and N sectors. See text for details.}
\label{tab:results}
\begin{tabular}{c c c c c c}
\hline \hline
Sector & Shell & $\sigma_{f}$ & kT [keV] & $v_{k, 1D}$ [km\,s$^{-1}$] & $\sqrt{C}$ \\ \hline
NE & 1& $0.21_{-0.02}^{+0.02}$& $6.9$& $532_{-45}^{+42}$& $1.08_{-0.01}^{+0.01}$ \\ 
 & 2& $0.24_{-0.03}^{+0.02}$& $7.4$& $631_{-67}^{+58}$& $1.11_{-0.02}^{+0.02}$ \\ 
 & 3& $0.09_{-0.08}^{+0.06}$& $7.2$& $228_{-211}^{+157}$& $1.01_{-0.01}^{+0.03}$ \\ 
 & 4& $0.4_{-0.19}^{+0.16}$& $7.2$& $1029_{-496}^{+409}$& $1.33_{-0.25}^{+0.41}$ \\  \hline
E & 1& $0.11_{-0.01}^{+0.01}$& $6.4$& $271_{-15}^{+22}$& $1.02_{-0.0}^{+0.0}$ \\ 
 & 2& $0.18_{-0.01}^{+0.01}$& $5.9$& $418_{-30}^{+32}$& $1.06_{-0.01}^{+0.01}$ \\ 
 & 3& $0.16_{-0.06}^{+0.04}$& $6.5$& $390_{-135}^{+90}$& $1.05_{-0.03}^{+0.02}$ \\ 
 & 4& $0.29_{-0.16}^{+0.09}$& $6.7$& $703_{-388}^{+217}$& $1.15_{-0.12}^{+0.12}$ \\  \hline
S & 1& $0.16_{-0.02}^{+0.01}$& $7.7$& $416_{-40}^{+29}$& $1.04_{-0.01}^{+0.01}$ \\ 
 & 2& $0.1_{-0.03}^{+0.03}$& $7.2$& $259_{-87}^{+87}$& $1.02_{-0.01}^{+0.01}$ \\ 
 & 3& $0.1_{-0.07}^{+0.05}$& $7.2$& $249_{-182}^{+139}$& $1.02_{-0.02}^{+0.02}$ \\ 
 & 4& $0.18_{-0.16}^{+0.12}$& $7.2$& $464_{-409}^{+301}$& $1.06_{-0.06}^{+0.11}$ \\  \hline
SW & 1& $0.13_{-0.02}^{+0.01}$& $7.7$& $331_{-49}^{+39}$& $1.03_{-0.01}^{+0.01}$ \\ 
 & 2& $0.0_{-0.0}^{+0.04}$& $7.2$& $12_{-12}^{+94}$& $1.0_{-0.0}^{+0.0}$ \\ 
 & 3& $0.14_{-0.1}^{+0.05}$& $7.2$& $365_{-252}^{+117}$& $1.04_{-0.03}^{+0.03}$ \\ 
 & 4& $0.28_{-0.15}^{+0.11}$& $7.2$& $721_{-392}^{+277}$& $1.15_{-0.12}^{+0.16}$ \\  \hline
W & 1& $0.06_{-0.01}^{+0.01}$& $7.7$& $169_{-29}^{+34}$& $1.01_{-0.0}^{+0.0}$ \\ 
 & 2& $0.05_{-0.01}^{+0.01}$& $6.9$& $131_{-27}^{+28}$& $1.0_{-0.0}^{+0.0}$ \\ 
 & 3& $0.18_{-0.02}^{+0.02}$& $5.8$& $401_{-42}^{+52}$& $1.05_{-0.01}^{+0.02}$ \\ 
 & 4& $0.02_{-0.02}^{+0.13}$& $6.6$& $48_{-48}^{+311}$& $1.0_{-0.0}^{+0.04}$ \\  \hline
NW & 1& $0.12_{-0.01}^{+0.01}$& $8.4$& $320_{-25}^{+33}$& $1.02_{-0.0}^{+0.01}$ \\ 
 & 2& $0.09_{-0.02}^{+0.01}$& $6.6$& $232_{-37}^{+35}$& $1.02_{-0.0}^{+0.01}$ \\ 
 & 3& $0.27_{-0.16}^{+0.08}$& $5.9$& $630_{-364}^{+177}$& $1.14_{-0.11}^{+0.1}$ \\  \hline
NWD & 1& $0.16_{-0.05}^{+0.04}$& $5.6$& $371_{-102}^{+92}$& $1.05_{-0.02}^{+0.03}$ \\  \hline
 N & 1& $0.08_{-0.01}^{+0.01}$& $7.6$& $219_{-22}^{+22}$& $1.01_{-0.0}^{+0.0}$ \\ 
 & 2& $0.23_{-0.03}^{+0.03}$& $8.0$& $615_{-73}^{+80}$& $1.09_{-0.02}^{+0.03}$ \\ 
 & 3& $0.24_{-0.05}^{+0.05}$& $7.6$& $620_{-122}^{+134}$& $1.1_{-0.04}^{+0.05}$ \\ 
 & 4& $0.25_{-0.19}^{+0.09}$& $7.4$& $641_{-485}^{+242}$& $1.11_{-0.11}^{+0.11}$ \\  \hline
Comb. & 1& $0.13_{-0.02}^{+0.02}$& $7.3$& $325_{-62}^{+60}$& $1.02_{-0.01}^{+0.01}$ \\ 
 & 2& $0.13_{-0.04}^{+0.04}$& $7.1$& $332_{-106}^{+107}$& $1.03_{-0.01}^{+0.03}$ \\ 
 & 3& $0.14_{-0.03}^{+0.03}$& $6.9$& $352_{-72}^{+74}$& $1.03_{-0.01}^{+0.02}$ \\ 
 & 4& $0.22_{-0.08}^{+0.06}$& $7.0$& $549_{-188}^{+158}$& $1.08_{-0.04}^{+0.07}$ \\  \hline
\end{tabular} \\

\end{table}}

Figure \ref{fig:modelvis} provides a visualization of the goodness of fit of the model for each shell in the N sector, showing posterior predictions for the distributions of surface brightness due to each model component, as well as their sum and the empirical brightness distribution of the data. Equivalent diagnostics for the other sectors are shown in Appendix \ref{apx:mvis}. 

As can be seen in Figures \ref{fig:modelvis} and \ref{fig:amodelvis}, our model provides a good description of the data in all shells. This gives us a post-facto justification for the assumption of log-normality for the (projected) cluster surface brightness distribution. Having verified that the model distributions match well to the data, we use the posterior distributions, obtained through MCMC, to evaluate the surface brightness fluctuations $\sigma_f$ in each shell. Because the posterior distributions are asymmetric, we report the modes and the 68.3 per cent highest posterior density intervals (HPDI).

In addition to the results for the individual sectors, we have also combined the results of all sectors except the NWD sector. For each of the 7 shells at a given radial range, we took samples $\sigma_{f, n}$ from the posterior distributions of $\sigma_f$, and then constructed a likelihood function  $\mathcal{L} = \rm{Norm}(\sigma_{f,n} | \mu_{\sigma_f}, \sigma_{\sigma_f})$, marginalizing over $N=5000$ number of samples, and then taking the logarithm and summing the log-likelihoods of the individual shells. The free parameters  $\mu_{\sigma_f}$ and $\sigma_{\sigma_f}$ represent the mean and the scatter in the surface brightness log-normal standard deviation $\sigma_f$. With this likelihood function, we used MCMC to find the posterior distributions for these two parameters, and calculated values for the combined sectors from the posterior distributions of $\mu_{\sigma_f}$. We also investigated the combined results for the combined quiescent arms (NE, S, SW, NWD and N sectors), and the arms along the E-W axis (E, W, and NW sectors), but found no statistically significant differences between the combined quiescent and combined E-W axis arms. 

In order to find the scaling factor between the two-dimensional surface brightness fluctuations and the 3-dimensional density fluctuations, we use a power spectrum analysis in several of the shells. We discuss power spectra and the relation between 2-dimensional and 3-dimensional fluctuations in more detail in Section \ref{sec:disc:3D}. In order to obtain the amplitude of the density fluctuations, we multiple $\sigma_f$ by a factor of 1.87.

We subsequently calculate the density bias $\sqrt{C}$ for the individual sectors as well as the combined results, using equation \ref{eq:clumpsig}. These biases in each shell along the sectors are shown in Figure \ref{fig:clump}.  While the intrinsic scatter between individual sectors below $1.6$\,$r_{2500}$ is somewhat large, in most individual shells we find the density bias to below $15\%$. For the combined results, we find the scatter to be better constrained, and about $3\%$ for inner radii. For the shells at $1.6--2.2$\,$r_{2500}$, the density bias is $8\%$, although given the uncertainties, this increase is only marginally significant. We also note that our analysis assumes that all observed density fluctuations can be attributed to the Perseus Cluster. Given the big solid angle that Perseus extends on the sky, it is likely that some extended background sources will contaminate our field of view and introduce additional fluctuations. It is therefore possible that the intrinsic density bias is somewhat lower than is measured here.

The largest-scale fluctuations are expected to dominate the overall contribution to the measured fluctuations, although to what degree requires knowledge of the slope of the spectrum. The shells are not equal in size, with the smallest inner shells extending about $10^\prime$ and the largest outer shells extending about $25^\prime$ in the azimuthal direction. This would suggest that the shells at larger radii are sensitive to somewhat larger-scale fluctuations, although no unambiguous trend of increasing bias with radius is observed. We refer to section \ref{sec:disc:3D} for further discussion of the length scales that influence our measurement. 

\cite{Zhuravleva2015} consider the clumping factor $C$ in terms of an integration over the density fluctuation power spectrum, in the inner $\sim 200\,$kpc region of Perseus. As can be seen from their results, the clumping factor is significantly lower if one puts an upper limit on the largest scale (in their case, $l = 100$ arcmin for Perseus, roughly comparable to the $25$ arcmin extent of the largest shells in our model). The $6-8\%$ clumping factor (and therefore a density bias $\sqrt{C}=3$--$4\%$) measured in the central region is very similar to our measurements at radii of $0.3-1.6$\,$r_{2500}$. Our results are also consistent with the clumping measurements of \cite{Simionescu2012} which used \textit{Suzaku} data, as in those data the clumping factor was measured to be close to 1 at smaller radii, only starting to increase at $\approx 0.5 r_{200} \approx 1.6\, r_{2500}$, corresponding to the outermost shells in our analysis. The results indicate that the amplitude of the clumping factor is modest, and unlikely to bias measurements of the ICM density, or cosmological measurements using the ICM density significantly. 

Secondly, we calculate the one-component turbulent velocity under the assumption that all the fluctuations that we observe are sourced by turbulent motions. In that case, we use equations \ref{eq:sbn} and \ref{eq:vturb}, using the standard deviation of the cluster log-normal model as representative fluctuations. Additionally, the turbulent spectrum depends on the sound speed, which is given by:
\begin{equation}
    c_s = \sqrt{ \frac{\gamma k T}{\mu m_p}}
\end{equation}
where $\gamma=5/3$ is the adiabatic index, $k$ is the Boltzmann constant, $T$ is the plasma temperature, $\mu=0.61$ is the mean particle weight, and $m_p$ is the proton mass. We extracted radial temperature profiles in various sectors, as described in appendix \ref{apx:temp}. The resulting turbulent velocities are shown in Figure \ref{fig:vturb}. 

Contrary to the density bias which represents a weighted sum of contributions over different scales, the turbulent velocity represents the velocity of the plasma at a particular length scale. Thus, the interpretation of what scale our measured velocities correspond to becomes more complex. Given that the measured velocity represents some type of summation of the velocities over the length scale range, we can state that the turbulent velocities that we measure represent an upper limit for length scales around $1^\prime$.

The full results for $\sigma_f$, the density bias and the turbulent velocity are shown in Table \ref{tab:results}. For each shell, we list $\sigma_{f}$ and the two derived quantities $v_{\rm 1D}$ (equation \ref{eq:vturb}) and $\sqrt{C}$ (equation \ref{eq:clumpsig}).

\section{Discussion}

\label{sec:discussion}

\subsection{Constraints on the presence of a high-density tail}

\begin{figure}
\centering
\vspace*{-5mm}
\includegraphics[width=0.99\linewidth]{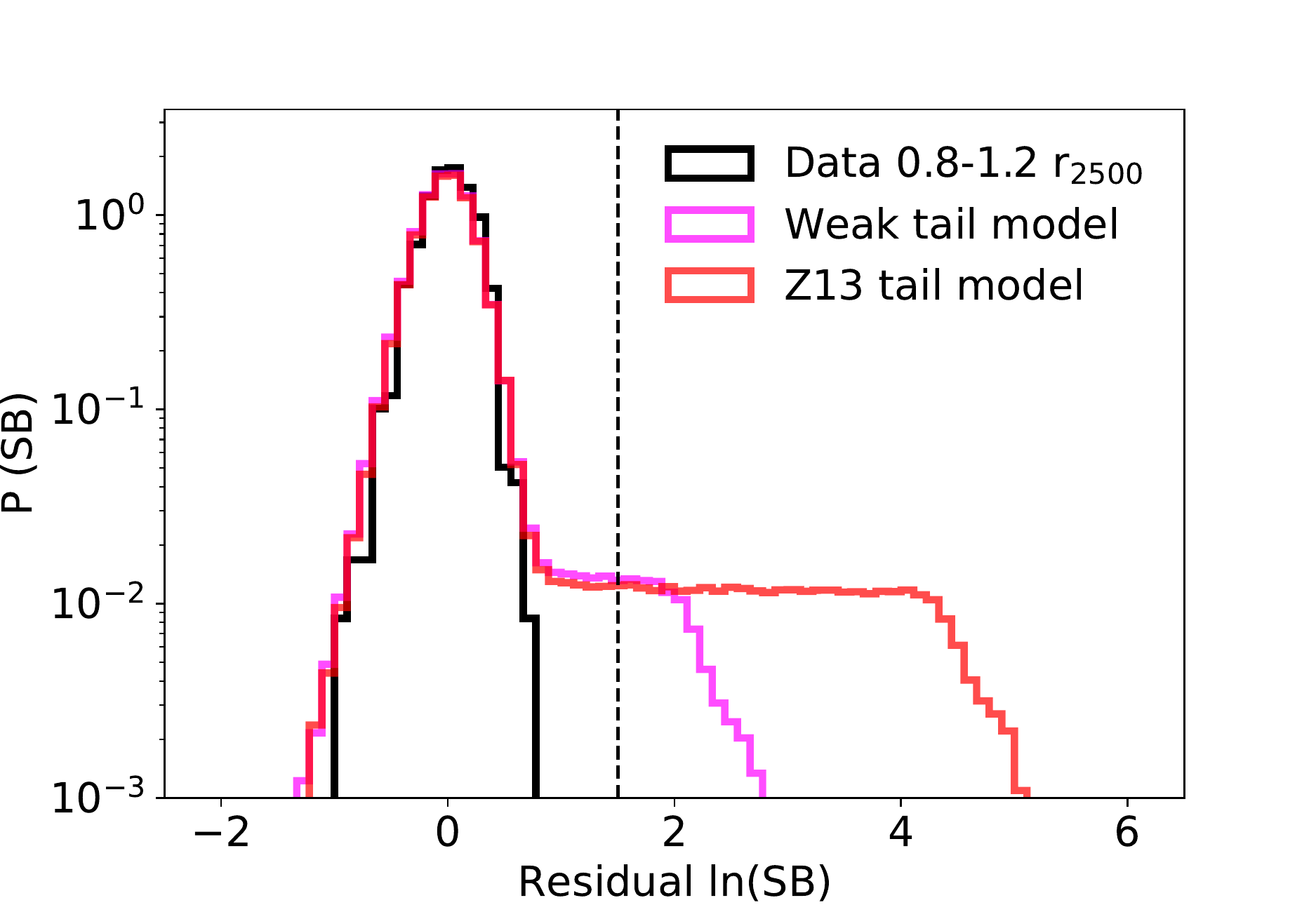} \\
\vspace*{-3mm}
 
\caption{Histogram of residual surface brightness predicted by 3000 sets of Monte-Carlo realisations of both tail models for the 5 shells in the 0.8--1.2\,$r_{2500}$ radial range in the NE, E, S, SW and N sectors. For comparison, the combined residual data of the 0.8--1.2\,$r_{2500}$ shells in those sectors are also plotted. The residuals were obtained by subtracting the mode of the distribution at each band in logspace. The dashed line indicates the threshold $\ln{\rm{SB}} > 1.5$ brighter than the mode of the distribution, where we quantitatively compare the data and predictions. }
\label{fig:lndev}
\end{figure}

{\renewcommand{\arraystretch}{1.1}

\begin{table}
\centering
\caption{Binomial test for the surface brightness data assuming two forms of a cluster emission model with a high-density tail. $f_{\rm tail}$  is defined as the fraction of Monte-Carlo samples that are $\ln{\rm{SB}} > 1.5$ brighter than the mode of the distribution. P(bins) is the probability that the tail model produces the number of tail bins seen in the data.}
\begin{tabular}{c c c c c l }
\hline \hline
$r_{2500}$ range  & N$_{\rm bins}$ &  Tail bins & Model  & $f_{\rm tail}$  & P(bins) \\ \hline
0.3--0.8 & 681  & 0 & Z13 &  0.0239  & $ 7.1 \times 10^{-8}$ \\
 & &  & weak & 0.0073 &  $ 6.6 \times 10^{-3}$ \\
0.8--1.2 &  1070  &0  & Z13 & 0.0360  & $ 9.3 \times 10^{-18}$\\
&&  & weak & 0.0100  & $ 2.0 \times 10^{-4}$   \\
1.2--1.6 & 1441 & 0 & Z13 &  0.0343  & $ 1.5 \times 10^{-22} $ \\
 &  &  & weak & 0.0059  & $ 1.7 \times 10^{-4}$\\
1.6--2.2 & 1413 & 0 & Z13 & 0.0173  & $ 2.2 \times 10^{-11}$ \\ 
 &  &  & weak & 0.0018 & $0.075$\\ \hline

\end{tabular}
\label{tab:binom}
\end{table}}

As discussed in Section \ref{sec:mod:eqs}, while basic theoretical arguments would suggest that the spectrum of density and, therefore, X-ray surface brightness fluctuations in the ICM is likely to be approximately log-normal in shape, hydrodynamical simulations have commonly predicted the presence of a high-density tail to this distribution, associated with the presence of cool, high surface brightness `clumps', \cite[e.g.][]{ Zhuravleva2013}. Whether these clumps are physically present or an artifact of the prescriptions used to model sub-grid physics, such as cooling, star formation and AGN feedback, has been a topic of debate. 

Our analysis allows us to place firm constraints on the presence of such bright, dense clumps in the Perseus Cluster. Figure \ref{fig:amodelvis} shows that the log normal model provides, by eye, an impressively good description of the Chandra data. In order to place quantitative constraints on the presence of high density tails to the distributions, we included a high-density tail in the cluster emission model component. Because the shape of such a high-density tail in real data is unknown, we defined both two tail models: the `Z13' model is based on the  probability density function presented in Figure 2 of \cite{Zhuravleva2013}. The height of the Z13 tail model was taken to be $0.25$ per cent of the peak of the cluster log-normal component. At this tail height, the intersection of the log-normal and tail components should occur at $\approx 3.462\,\sigma_f$. The extent of the tail was then set at an order of magnitude in density beyond this intersection, which by equation \ref{eq:sbn} equates to two orders of magnitude for surface brightness. The `weak' tail model is set at the same height and thus intersects the log-normal component at the same point, but only extends half an order of magnitude in density (thus one order of magnitude in surface brightness). 

Using these models, we generated 3000 sets of surface brightness Monte-Carlo realisations in several shells in the same manner as the realizations shown in Figure \ref{fig:modelvis}. For efficiency, we used the MCMC parameter samples obtained from the non-tail model to describe the log normal cluster emission components.\footnote{Since the tail is a relatively small perturbation to the log-normal component, re-running the full MCMC analysis including the tail should change the posterior distributions for the ICM model parameters by a negligible amount.} We obtained residual surface brightness distributions by subtracting the mode of the distribution in each band, and then combined all bins from the 5 sectors that use the same radial ranges (NE, E, S, SW and N sectors) into a single distribution. Figure \ref{fig:lndev} shows the combined predicted residual distributions for both tail models compared with the data for the 5 shells at 0.8--1.2\,$r_{2500}$.

We computed the binomial probability for the number of bins in the data at surface brightness $\ln{\rm{SB}} > 1.5$  brighter than the mode of the distribution (i.e. unambiguously in the tail) with the predictions of both tail models. Table \ref{tab:binom} shows the probabilities of the tail models producing the observed number of bins in this region, which is zero bins in all cases. Because there is stochasticity to the predictions (as a result of the Monte-Carlo method), we report the highest probability after a minimum of 1500 realizations of the data set as the most conservative estimate. For the Z13 tail model, the probability at all radial ranges is smaller than $7.1 \times 10^{-8}$, which argues strongly against the presence of a tail such as the one seen in \cite{Zhuravleva2013}. For the weak tail model, the probability remain below $6.6 \times 10^{-3}$ for the inner 3 radial ranges. For the 1.6--2.2\,$r_{2500}$ shells, the background is strong enough that the weak tail model barely extends beyond the cutoff. As a result we are not able to completely rule out the presence of a weak tail model at this radius. The binomial probability with the predictions of the non-tailed model is $>95\%$ at all radial ranges. 

The tails observed in \cite{Zhuravleva2013} were seen at radii $>r_{500}$, approximately corresponding to the outer edge of our analysis. Although our analysis strongly rules out these tails in the Perseus cluster at $0.3$--$2.2$\,$r_{2500}$, it is possible that such tails still might exist in real data at larger radii.

We note that clumps dense and small enough to be resolved by \textit{Chandra} would have been identified as point sources and excluded from the analysis. However, the excellent agreement in Figure \ref{fig:agnpred} between the modeled and observed number of AGN indicates that the number of bright surface brightness clumps misidentified in such manner is likely to be negligible. 

\subsection{Comparison to power spectra}
\label{sec:disc:3D}

\begin{figure}
\centering
\includegraphics[width=0.99\linewidth]{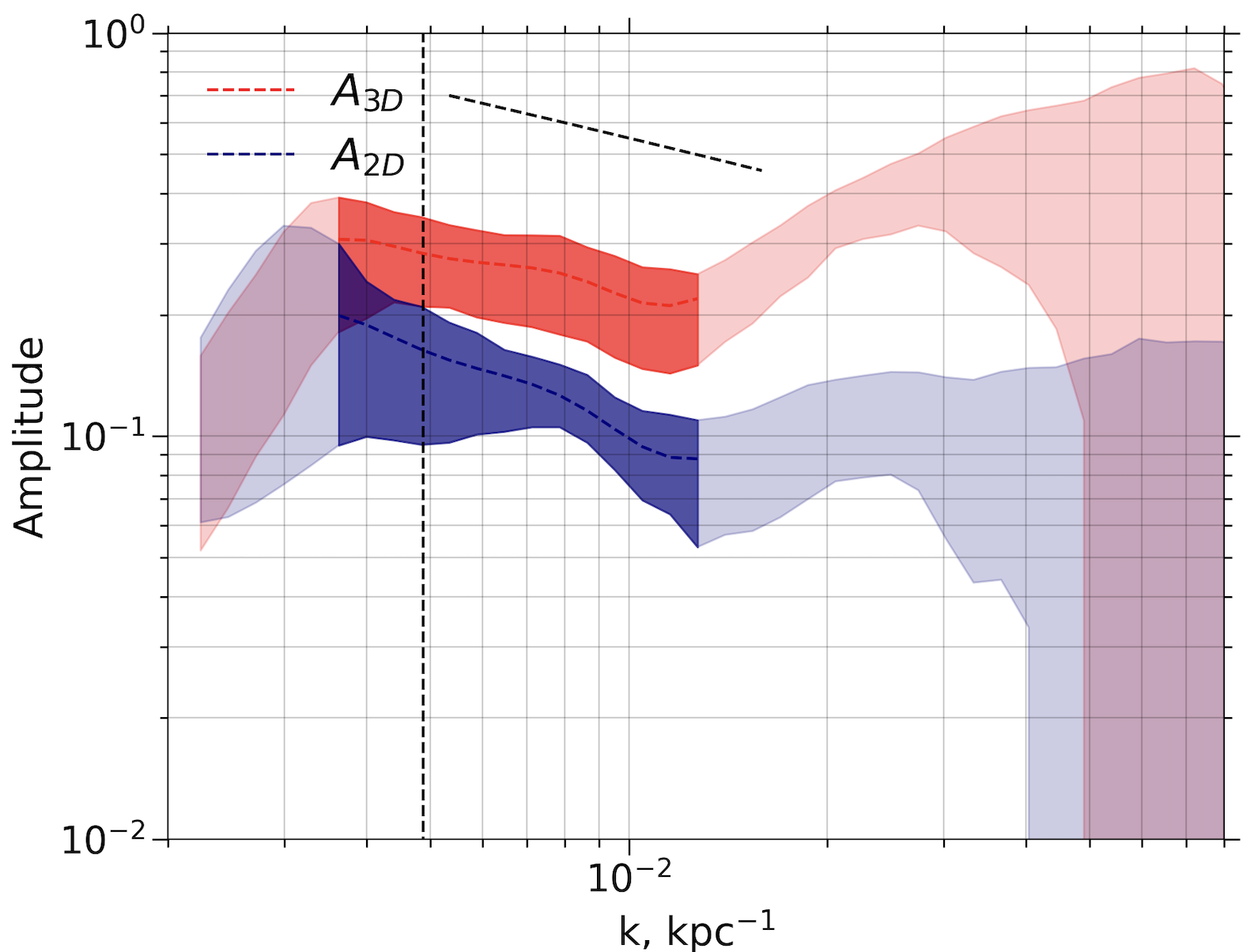} \\
\caption{Power spectrum for the inner shell in the S sector. The blue curve shows the amplitude of 2D surface brightness fluctuations A$_{2D}$, the red curve shows the amplitude of 3D density fluctuations A$_{3D}$. The dashed vertical line indicates the effective scale at which A$_{2D}$ corresponds to our measurement for this shell. The dashed diagonal line indicates the slope of a Kolmogorov turbulent spectrum. }
\label{fig:PSpec}
\end{figure}

As briefly mentioned in Section \ref{sec:results}, we have conducted a comparison with power spectra for a small subset of shells in which the number of counts were sufficient to get meaningful measurements at relevant scales. The power spectrum analysis aids interpretation of our results in two key ways. Firstly, our measurement represents a weighted sum of contributions over a range of scales between the size of the spatial bin (1 by 1 arcmin) and the size of the shell (of order 15 arcmin). At the same time, fluctuations along the line of sight will also contribute to the measurement. Because of the positive correlation between fluctuation amplitude and scale, larger scales are expected to dominate the overall measurement. By comparing with the power spectrum and identifying the scale where the power spectrum analysis matches our measurement, we can thus infer an `effective scale' that our measurements correspond to. Secondly, because the power spectrum method includes a deprojection analysis, it allows us to estimate the amplitude of the 3-dimensional fluctuations from the measured 2-dimensional fluctuations. These projection effects were not directly accounted for in our forward-model, primarily because estimating the contribution to the amplitude of fluctuations of projected emission at larger radius would require making a \textit{a priori} assumptions about the scale-dependent nature of those fluctuations. In general, projection has the effect of suppressing the observed amplitude of the fluctuations, as the contribution of random fluctuations will be averaged over the line of sight. For smaller-scale fluctuations, this suppression will be stronger as the projected line of sight through the cluster will extend across a larger number of fluctuations. 

We have followed the power spectrum method as described in \cite{Zhuravleva2015}, and the deprojection method laid out in \cite{Churazov2012}, and generated power spectra in the $0.3$--$0.8$\,$r_{2500}$ shells of the NE, E and S sectors. We also attempted to generate a power spectrum of the $0.3$--$0.8$\,$r_{2500}$ shell in the N sector, but this sector did not contain sufficient counts and yielded a very noisy spectrum. For the deprojection procedure, we assumed a spherically symmetric $\beta$-model for each shell (by using a separate $\beta$-model for each individual shell). In principle, large-scale perturbations can be included in the $\beta$-model, which would result in a shallower power spectrum because the large-scale perturbations are divided out, rather than being included in the power spectrum. Because it is difficult to know which perturbations are sourced by turbulence, and which perturbations are caused by other processes, such as large-scale disturbances of the gravitational potential (as might be the case along the E-W axis, which is not dynamically relaxed), we have chosen to use the spherically symmetric $\beta$-model as the most conservative option, essentially including all perturbations in the data in our estimates for the clumping factor and the turbulent velocity. This means that the inferred turbulent velocities should be seen as upper limits, under the assumption that all perturbations are sourced by turbulence. 

The power spectrum for the inner shell of the S sector can be seen in Figure \ref{fig:PSpec}, showing both the surface brightness power spectrum in terms of the amplitude A$_{2D}$, and the density power spectrum in terms of A$_{3D}$. For the 3 shells we have investigated, we find that the wave number at which the A$_{2D}$ amplitude equals the value of $\sigma_f$ in our analysis is $k \approx 0.005 $ kpc$^{-1}$ in all 3 shells, corresponding to an effective scale of 9.2 arcmin. This is consistent with the idea that the largest coherent scales spanned by our measurements (approximately 12 arcmin for the shells considered here) dominate the overall measurements.

At the approximate effective wave number $k=0.005$ kpc$^{-1}$, we determine the ratio between the amplitude of the 2D and 3D spectra to estimate how much projection effects suppress the measurements. We find that at the effective scale, the ratio $A_{3D}/A_{2D}$ equals 2.3, 1.6 and 1.7 for the inner shells in the NE, E and S sectors respectively. We use the average of these 3 values ($1.87$) to convert our measurements to the amplitude of density fluctuations in Section \ref{sec:results}. Because of strong Poisson noise, we could only generate power spectra for shells at the smallest radii. We make the assumption that this ratio between 2D and 3D fluctuations holds for shells at larger radii as well. Although the projected line of sight through the cluster becomes smaller as one moves to larger radii, the 3-dimensional emission regions contributing the most to the observed surface brightness along a given line of sight are the regions that are of similar brightness. Therefore, the effective length of the line of sight, i.e. the length of the region that most of the photons come from, should not vary dramatically as one move to larger radii. Based on the ratios observed in the 3 sectors (2.3, 1.6, and 1.7), we estimate that there is a systematic uncertainty of order $25\%$ in the calculated amplitude of the density fluctuations. As noted above, these values and the associated systematic uncertainty are only valid for the spherically symmetric $\beta$-model. For $\beta$-models that include large-scale perturbations, the ratio $A_{3D}/A_{2D}$ at the effective scale would decrease, and the resulting clumping factors and turbulent velocities would be lower.


\subsection{Turbulence in cluster simulations}

\begin{figure}
\centering
\includegraphics[width=0.99\linewidth]{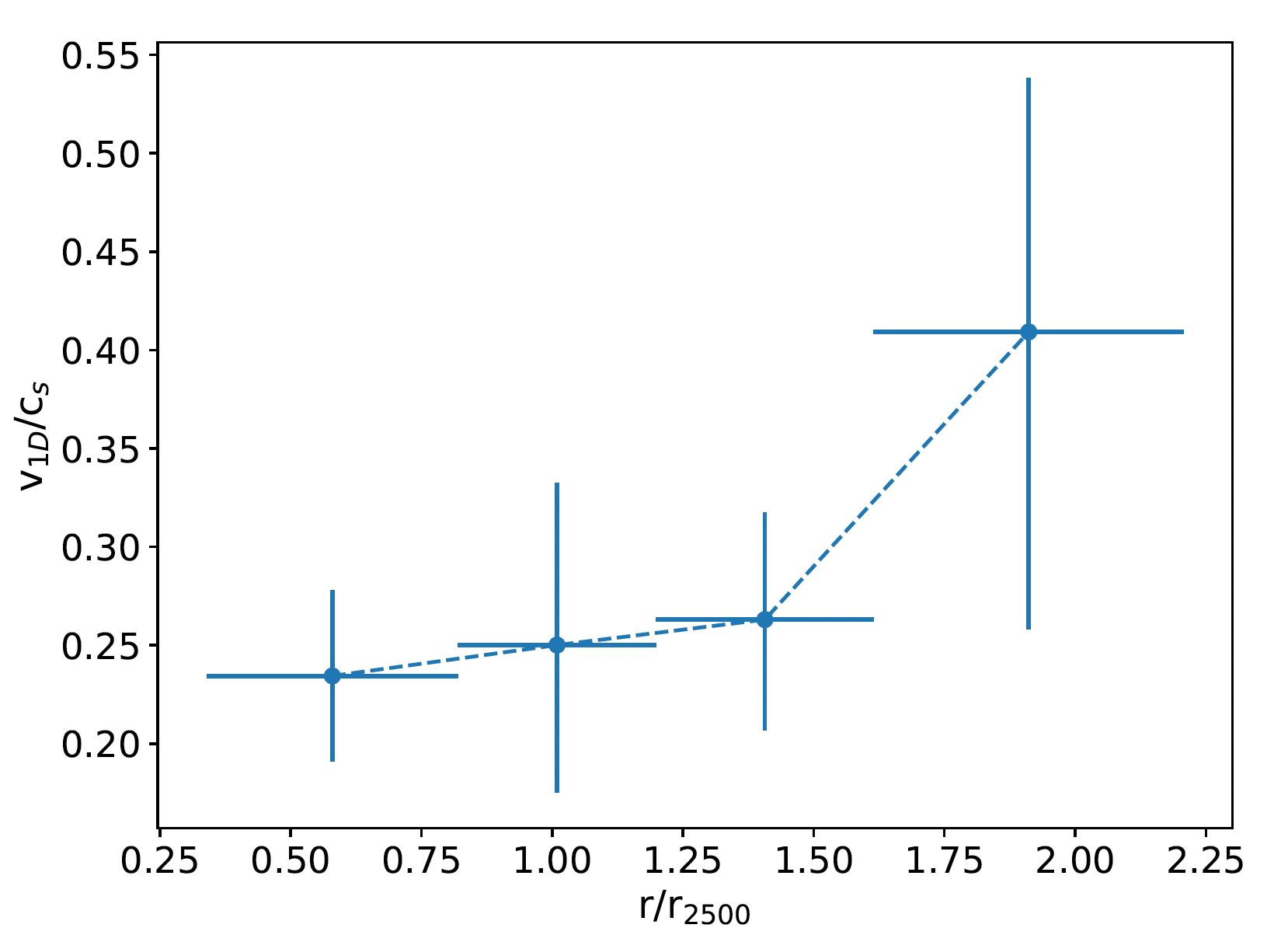} \\
\caption{The ratio of the inferred turbulent velocity to the sound speed as a function of radius, for the combined results.} 
\label{fig:csfrac}
\end{figure}

We have used the calculated turbulent velocities and compared them with the results of numerical simulations of the ICM. Of particular interest is the ratio of turbulent velocity to the sound speed, $v_{1D}/c_s$, as this gives an indication of the level of non-thermal pressure support that is caused by subsonic motions. The ratio of the energy density in turbulence to the thermal energy can be written as

\begin{equation}
    f_{\rm nth} = \frac{3 \gamma}{2} \left( \frac{v_{1D}}{c_s} \right)^2 .
\end{equation}

In the simulations described in \cite{Lau2009}, the ratio $v_{1D}/c_s$ is shown to range from $\approx 0.125$ to $0.3$ for radii between $0.2$--$1.0r_{500}$ in relaxed clusters, which will contribute $\approx 5-15\%$ to the total pressure support. Figure \ref{fig:csfrac} shows this fraction for our inferred turbulent velocities, for the combined sectors from the results in Table \ref{tab:results}. Although we also computed the results for the combined quiescent arms (NE, S, SW, NWD and N sectors) and the combined E-W axis arms (E, W, and NW sectors), we did not find any statistically significant differences between these two and therefore did not include them in the figure. Overall, we find marginal evidence for $v_{1D}/c_s$ increasing with radius: a power law fit the combined turbulent velocity profile results in a power law index of $0.18\pm 0.17$.

Our measured fraction is slightly higher than the predictions of \cite{Lau2009}. This is unsurprising, given our assumption that the observed X-ray surface brightness fluctuations are entirely sourced by turbulent motions. We We also note that on top of the statistical errors shown in Figure \ref{fig:csfrac}, there is a 25 per cent systematic uncertainty introduced by the conversion from surface brightness to density fluctuations (Section \ref{sec:disc:3D}). Thus, we can conclude that our results on the amplitude of turbulent velocities are broadly consistent with the predictions from hydrodynamical simulations.

\subsection{Current and Future Observations}

Although the current work has focused exclusively on \textit{Chandra} data, we note that Perseus has also been observed extensively with the \textit{XMM-Newton} telescope \citep[see e.g.][]{Simionescu2012, Sanders2020}, with good coverage of the cluster up to $\approx 1.5$\,$r_{2500}$. Although an analysis of those data is beyond the scope of this work, we note that the higher effective area of \textit{XMM} and the comparatively longer exposure times in certain regions at larger radii could in principle improve the uncertainties on our measurements. On the other hand, the CCD detectors of \textit{XMM} have some disadvantages as well, such as the fact that the particle background is more prominent and more variable, and and the larger PSF, which makes the removal and modeling of residual point sources more challenging. One area in which we envision \textit{XMM} data could be useful for future work is in providing more $>3.5$ keV photons, which we could use for a comparison between the soft (0.6--3.5\,keV) and hard (3.5--7.0 keV) X-ray bands, such that we could infer more about the thermodynamic processes that source the observed fluctuations \citep{Zhuravleva2015}.

In terms of future missions, both the upcoming \textit{XRISM} mission (planned launch February 2023) as well as \textit{ATHENA} in the next decade will shed more light on dynamical processes in galaxy clusters. Both of these missions will carry X-ray calorimeter instruments, allowing for eV-scale energy resolution and accurate radial velocity measurements of gas motions. With the sensitivity of \textit{XRISM}, radial velocity measurements should be possible out to radii of at least 0.4\,$r_{2500}$ in the Perseus Cluster (in small 3x3 arcmin$^2$ regions).  \citep{XRISM2020}. The 2016 \textit{Hitomi} mission made such a measurement in the core of Perseus, and found a line-of-sight turbulent velocity of $(164 \pm 10) \,\rm km\,s^{-1}$ at radii of 30-60\,kpc ($\approx 0.05-0.1\,r_{2500}$) \citep{Hitomi2016}. The turbulent velocity up to 100\,kpc away from the central AGN was measured to be constant at $\approx 100$km\,s$^{-1}$ \citep{Hitomi2018}.

Given that \textit{XRISM} will measure the gas velocity from spectral lines, it should not suffer the same suppression due to projection that occurs in our data. Patches of cluster gas along the line of sight will all contribute to Doppler-broadening of the total spectral line in the projected 2-dimensional region - which is fundamentally different from our method, where the scalar property of emissivity is averaged along the line of sight. The gas velocities observed with \textit{XRISM} in the Perseus Cluster, if only for a few small 3x3 arcmin$^2$ regions,  will thus still provide an important cross-check on our results presented in Table \ref{tab:results}.  

A less straightforward question is the effective scale at which \textit{XRISM} will measure the turbulent velocity. Although the beam-size of \textit{XRISM} is of order 1 arcmin, to some extent scales along the line of sight through the cluster might increase the effective scale at which velocities are measured. Assuming that the effective scale observed by \textit{XRISM} is 1 arcmin, and taking the effective scale of 9.2 arcmin from our data, and assuming and the power spectrum follows a Kolmogorov slope, we would expect that the velocities measured by \textit{XRISM} would be lower by a factor of $(9.2/1)^{1/3} \approx 2$ compared to what we measure. 


\section{Conclusions}
\label{sec:conclusion}
Using a large set of \textit{Chandra} observations of the Perseus Cluster, we have conducted a rigorous study of projected gas fluctuations in between radii of $0.3$ and $2.2\,r_{2500}$ along 8 different sectors. Our findings show that the density bias $\sqrt{C}$ averages to around $3_{-1}^{+2}$ per cent between radii of $0.3$--$1.6\,r_{2500}$. In the range of $1.6$--$2.2\,r_{2500}$ the density bias marginally increases to $8_{-4}^{+7}$ per cent. Because of the uncertainty in estimating 3-dimensional density fluctuations from 2-dimension surface brightness fluctuations, our measurements contain an additional $\approx 25$ per cent systematic error in the amplitude of $\sigma_{\rho}$, which for $\sqrt{C}$ translates to an additional uncertainty of $\approx 1$--$2$ per cent at $0.3$--$1.6\,r_{2500}$, and $\approx 5$--$6$ per cent at $1.6$--$2.2\,r_{2500}$. Overall, our findings show that the density bias will have a small effect on the measured gas mass in a cluster. 

At the radii $0.3$--$1.6\,r_{2500}$, we find an average turbulent velocity of $340_{-80}^{+80}$\,km\,s$^{-1}$, while the average turbulent velocity at $1.6$--$2.2\,r_{2500}$ is measured to be $549_{-188}^{+159}\,{\rm km}\,{\rm s}^{-1}$. Just as for $\sqrt{C}$, our measurements carry an additional $\approx 25$ per cent systematic uncertainty. Importantly, the turbulent velocity is calculated under the assumption that all the observed fluctuations are sourced by turbulence. This may not hold true in shells where large-scale fluctuations are caused by disturbance of the gravitational potential. As such, our measurements of the turbulent velocity might be seen as an upper limit.

The modeling done in this paper provides a blueprint for the analysis of gas fluctuations that can be extended and improved in a number of ways. By combining \textit{Chandra} and \textit{XMM-Newton} observations of the same system, the spatial resolution of the former can be used together with the higher sensitivity of the latter. Additionally, the joint analysis of a soft and hard energy band could reveal more about the nature of the gas fluctuations. Finally, it would be valuable to apply this analysis to other clusters in order to investigate the validity of these results for similar quiescent clusters.

\section*{Data Availability}
The \textit{Chandra} X-ray data used in this work (see Table \ref{tab:obsids}) are publicly available from the Chandra Data Archive (CDA) at \url{https://cxc.harvard.edu/cda/}. The ROSAT observations are publicly available from the ROSAT data archive \url{https://heasarc.gsfc.nasa.gov/docs/rosat/archive_access.html}. The X-ray surface brightness forward-modeling code can be found at \url{https://github.com/martijndevries/SB-forwardmodel}. All data shown in figures and tables can be obtained in digital form from \url{https://github.com/martijndevries/perseus-perturbations-paper}.

\section*{Acknowledgements}
This research has made use of data obtained from the Chandra Data Archive and the Chandra Source Catalog, and software provided by the Chandra X-ray Center (CXC) in the application package CIAO. Support for this work was provided by the National Aeronautics and Space Administration through Chandra Award number ****, issued by the Chandra X-ray Center, which is operated by the Smithsonian Astrophysical Observatory for and on behalf of the National Aeronautics Space Administration under contract NAS8-03060. This work was in part supported by the U.S. Department of Energy under contract number DE-AC02-76SF00515. 

\bibliographystyle{mnras}
\bibliography{Perseus_P}

\begin{thebibliography}{}
\makeatletter
\relax
\def\mn@urlcharsother{\let\do\@makeother \do\$\do\&\do\#\do\^\do\_\do\%\do\~}
\def\mn@doi{\begingroup\mn@urlcharsother \@ifnextchar [ {\mn@doi@}
  {\mn@doi@[]}}
\def\mn@doi@[#1]#2{\def\@tempa{#1}\ifx\@tempa\@empty \href
  {http://dx.doi.org/#2} {doi:#2}\else \href {http://dx.doi.org/#2} {#1}\fi
  \endgroup}
\def\mn@eprint#1#2{\mn@eprint@#1:#2::\@nil}
\def\mn@eprint@arXiv#1{\href {http://arxiv.org/abs/#1} {{\tt arXiv:#1}}}
\def\mn@eprint@dblp#1{\href {http://dblp.uni-trier.de/rec/bibtex/#1.xml}
  {dblp:#1}}
\def\mn@eprint@#1:#2:#3:#4\@nil{\def\@tempa {#1}\def\@tempb {#2}\def\@tempc
  {#3}\ifx \@tempc \@empty \let \@tempc \@tempb \let \@tempb \@tempa \fi \ifx
  \@tempb \@empty \def\@tempb {arXiv}\fi \@ifundefined
  {mn@eprint@\@tempb}{\@tempb:\@tempc}{\expandafter \expandafter \csname
  mn@eprint@\@tempb\endcsname \expandafter{\@tempc}}}

\bibitem[\protect\citeauthoryear{{Allen}, {Schmidt}, {Fabian}  \&
  {Ebeling}}{{Allen} et~al.}{2003}]{Allen2003}
{Allen} S.~W.,  {Schmidt} R.~W.,  {Fabian} A.~C.,   {Ebeling} H.,  2003,
  \mn@doi [\mnras] {10.1046/j.1365-8711.2003.06550.x}, \href
  {https://ui.adsabs.harvard.edu/abs/2003MNRAS.342..287A} {342, 287}

\bibitem[\protect\citeauthoryear{{Allen}, {Schmidt}, {Ebeling}, {Fabian}  \&
  {van Speybroeck}}{{Allen} et~al.}{2004}]{Allen2004}
{Allen} S.~W.,  {Schmidt} R.~W.,  {Ebeling} H.,  {Fabian} A.~C.,   {van
  Speybroeck} L.,  2004, \mn@doi [\mnras] {10.1111/j.1365-2966.2004.08080.x},
  \href {https://ui.adsabs.harvard.edu/abs/2004MNRAS.353..457A} {353, 457}

\bibitem[\protect\citeauthoryear{{Allen}, {Rapetti}, {Schmidt}, {Ebeling},
  {Morris}  \& {Fabian}}{{Allen} et~al.}{2008}]{Allen2008}
{Allen} S.~W.,  {Rapetti} D.~A.,  {Schmidt} R.~W.,  {Ebeling} H.,  {Morris}
  R.~G.,   {Fabian} A.~C.,  2008, \mn@doi [\mnras]
  {10.1111/j.1365-2966.2007.12610.x}, \href
  {https://ui.adsabs.harvard.edu/abs/2008MNRAS.383..879A} {383, 879}

\bibitem[\protect\citeauthoryear{{Allen}, {Evrard}  \& {Mantz}}{{Allen}
  et~al.}{2011}]{Allen2011}
{Allen} S.~W.,  {Evrard} A.~E.,   {Mantz} A.~B.,  2011, \mn@doi [\araa]
  {10.1146/annurev-astro-081710-102514}, \href
  {https://ui.adsabs.harvard.edu/abs/2011ARA&A..49..409A} {49, 409}

\bibitem[\protect\citeauthoryear{{Angelinelli}, {Ettori}, {Vazza}  \&
  {Jones}}{{Angelinelli} et~al.}{2021}]{Angelinelli2021}
{Angelinelli} M.,  {Ettori} S.,  {Vazza} F.,   {Jones} T.~W.,  2021, \mn@doi
  [\aap] {10.1051/0004-6361/202140471}, \href
  {https://ui.adsabs.harvard.edu/abs/2021A&A...653A.171A} {653, A171}

\bibitem[\protect\citeauthoryear{{Ar{\'e}valo}, {Churazov}, {Zhuravleva},
  {Forman}  \& {Jones}}{{Ar{\'e}valo} et~al.}{2016}]{Arevalo2016}
{Ar{\'e}valo} P.,  {Churazov} E.,  {Zhuravleva} I.,  {Forman} W.~R.,   {Jones}
  C.,  2016, \mn@doi [\apj] {10.3847/0004-637X/818/1/14}, \href
  {https://ui.adsabs.harvard.edu/abs/2016ApJ...818...14A} {818, 14}

\bibitem[\protect\citeauthoryear{{Bluem} et~al.,}{{Bluem}
  et~al.}{2022}]{Bluem2022}
{Bluem} J.,  et~al., 2022, arXiv e-prints, \href
  {https://ui.adsabs.harvard.edu/abs/2022arXiv220802477B} {p. arXiv:2208.02477}

\bibitem[\protect\citeauthoryear{{Churazov} et~al.,}{{Churazov}
  et~al.}{2012}]{Churazov2012}
{Churazov} E.,  et~al., 2012, \mn@doi [\mnras]
  {10.1111/j.1365-2966.2011.20372.x}, \href
  {https://ui.adsabs.harvard.edu/abs/2012MNRAS.421.1123C} {421, 1123}

\bibitem[\protect\citeauthoryear{{Churazov}, {Arevalo}, {Forman}, {Jones},
  {Schekochihin}, {Vikhlinin}  \& {Zhuravleva}}{{Churazov}
  et~al.}{2016}]{Churazov2016}
{Churazov} E.,  {Arevalo} P.,  {Forman} W.,  {Jones} C.,  {Schekochihin} A.,
  {Vikhlinin} A.,   {Zhuravleva} I.,  2016, \mn@doi [\mnras]
  {10.1093/mnras/stw2044}, \href
  {https://ui.adsabs.harvard.edu/abs/2016MNRAS.463.1057C} {463, 1057}

\bibitem[\protect\citeauthoryear{{Dunn} \& {Fabian}}{{Dunn} \&
  {Fabian}}{2006}]{Dunn2006}
{Dunn} R.~J.~H.,  {Fabian} A.~C.,  2006, \mn@doi [\mnras]
  {10.1111/j.1365-2966.2006.11080.x}, \href
  {https://ui.adsabs.harvard.edu/abs/2006MNRAS.373..959D} {373, 959}

\bibitem[\protect\citeauthoryear{{Eckert}, {Roncarelli}, {Ettori}, {Molendi},
  {Vazza}, {Gastaldello}  \& {Rossetti}}{{Eckert} et~al.}{2015}]{Eckert2015}
{Eckert} D.,  {Roncarelli} M.,  {Ettori} S.,  {Molendi} S.,  {Vazza} F.,
  {Gastaldello} F.,   {Rossetti} M.,  2015, \mn@doi [\mnras]
  {10.1093/mnras/stu2590}, \href
  {https://ui.adsabs.harvard.edu/abs/2015MNRAS.447.2198E} {447, 2198}

\bibitem[\protect\citeauthoryear{{Gaspari}, {Churazov}, {Nagai}, {Lau}  \&
  {Zhuravleva}}{{Gaspari} et~al.}{2014}]{Gaspari2014}
{Gaspari} M.,  {Churazov} E.,  {Nagai} D.,  {Lau} E.~T.,   {Zhuravleva} I.,
  2014, \mn@doi [\aap] {10.1051/0004-6361/201424043}, \href
  {https://ui.adsabs.harvard.edu/abs/2014A&A...569A..67G} {569, A67}

\bibitem[\protect\citeauthoryear{{Goodman} \& {Weare}}{{Goodman} \&
  {Weare}}{2010}]{Goodman2010}
{Goodman} J.,  {Weare} J.,  2010, \mn@doi [Communications in Applied
  Mathematics and Computational Science] {10.2140/camcos.2010.5.65}, \href
  {https://ui.adsabs.harvard.edu/abs/2010CAMCS...5...65G} {5, 65}

\bibitem[\protect\citeauthoryear{{Hickox} \& {Markevitch}}{{Hickox} \&
  {Markevitch}}{2006}]{Hickox2005}
{Hickox} R.~C.,  {Markevitch} M.,  2006, \mn@doi [\apj] {10.1086/504070}, \href
  {https://ui.adsabs.harvard.edu/abs/2006ApJ...645...95H} {645, 95}

\bibitem[\protect\citeauthoryear{{Hitomi Collaboration} et~al.,}{{Hitomi
  Collaboration} et~al.}{2016}]{Hitomi2016}
{Hitomi Collaboration} et~al., 2016, \mn@doi [\nat] {10.1038/nature18627},
  \href {https://ui.adsabs.harvard.edu/abs/2016Natur.535..117H} {535, 117}

\bibitem[\protect\citeauthoryear{{Hitomi Collaboration} et~al.,}{{Hitomi
  Collaboration} et~al.}{2018}]{Hitomi2018}
{Hitomi Collaboration} et~al., 2018, \mn@doi [\pasj] {10.1093/pasj/psx138},
  \href {https://ui.adsabs.harvard.edu/abs/2018PASJ...70....9H} {70, 9}

\bibitem[\protect\citeauthoryear{{Kawahara}, {Suto}, {Kitayama}, {Sasaki},
  {Shimizu}, {Rasia}  \& {Dolag}}{{Kawahara} et~al.}{2007}]{Kawahara2007}
{Kawahara} H.,  {Suto} Y.,  {Kitayama} T.,  {Sasaki} S.,  {Shimizu} M.,
  {Rasia} E.,   {Dolag} K.,  2007, \mn@doi [\apj] {10.1086/512231}, \href
  {https://ui.adsabs.harvard.edu/abs/2007ApJ...659..257K} {659, 257}

\bibitem[\protect\citeauthoryear{{Kawahara}, {Reese}, {Kitayama}, {Sasaki}  \&
  {Suto}}{{Kawahara} et~al.}{2008}]{Kawahara2008}
{Kawahara} H.,  {Reese} E.~D.,  {Kitayama} T.,  {Sasaki} S.,   {Suto} Y.,
  2008, \mn@doi [\apj] {10.1086/591930}, \href
  {https://ui.adsabs.harvard.edu/abs/2008ApJ...687..936K} {687, 936}

\bibitem[\protect\citeauthoryear{{Khedekar}, {Churazov}, {Kravtsov},
  {Zhuravleva}, {Lau}, {Nagai}  \& {Sunyaev}}{{Khedekar}
  et~al.}{2013}]{Khedekar2013}
{Khedekar} S.,  {Churazov} E.,  {Kravtsov} A.,  {Zhuravleva} I.,  {Lau} E.~T.,
  {Nagai} D.,   {Sunyaev} R.,  2013, \mn@doi [\mnras] {10.1093/mnras/stt224},
  \href {https://ui.adsabs.harvard.edu/abs/2013MNRAS.431..954K} {431, 954}

\bibitem[\protect\citeauthoryear{{Lau}, {Kravtsov}  \& {Nagai}}{{Lau}
  et~al.}{2009}]{Lau2009}
{Lau} E.~T.,  {Kravtsov} A.~V.,   {Nagai} D.,  2009, \mn@doi [\apj]
  {10.1088/0004-637X/705/2/1129}, \href
  {https://ui.adsabs.harvard.edu/abs/2009ApJ...705.1129L} {705, 1129}

\bibitem[\protect\citeauthoryear{{Liu}, {Fabian}, {Pinto}, {Russell}, {Sanders}
   \& {McNamara}}{{Liu} et~al.}{2021}]{Liu2021}
{Liu} H.,  {Fabian} A.~C.,  {Pinto} C.,  {Russell} H.~R.,  {Sanders} J.~S.,
  {McNamara} B.~R.,  2021, \mn@doi [\mnras] {10.1093/mnras/stab1372}, \href
  {https://ui.adsabs.harvard.edu/abs/2021MNRAS.505.1589L} {505, 1589}

\bibitem[\protect\citeauthoryear{{Mantz}, {Allen}, {Morris}, {Rapetti},
  {Applegate}, {Kelly}, {von der Linden}  \& {Schmidt}}{{Mantz}
  et~al.}{2014}]{Mantz2014}
{Mantz} A.~B.,  {Allen} S.~W.,  {Morris} R.~G.,  {Rapetti} D.~A.,  {Applegate}
  D.~E.,  {Kelly} P.~L.,  {von der Linden} A.,   {Schmidt} R.~W.,  2014,
  \mn@doi [\mnras] {10.1093/mnras/stu368}, \href
  {https://ui.adsabs.harvard.edu/abs/2014MNRAS.440.2077M} {440, 2077}

\bibitem[\protect\citeauthoryear{{Mantz}, {Allen}, {Morris}, {Schmidt}, {von
  der Linden}  \& {Urban}}{{Mantz} et~al.}{2015}]{Mantz2015}
{Mantz} A.~B.,  {Allen} S.~W.,  {Morris} R.~G.,  {Schmidt} R.~W.,  {von der
  Linden} A.,   {Urban} O.,  2015, \mn@doi [\mnras] {10.1093/mnras/stv219},
  \href {https://ui.adsabs.harvard.edu/abs/2015MNRAS.449..199M} {449, 199}

\bibitem[\protect\citeauthoryear{{Mantz} et~al.,}{{Mantz}
  et~al.}{2022}]{Mantz2022}
{Mantz} A.~B.,  et~al., 2022, \mn@doi [\mnras] {10.1093/mnras/stab3390}, \href
  {https://ui.adsabs.harvard.edu/abs/2022MNRAS.510..131M} {510, 131}

\bibitem[\protect\citeauthoryear{{Mirakhor} \& {Walker}}{{Mirakhor} \&
  {Walker}}{2021}]{Mirakhor2021}
{Mirakhor} M.~S.,  {Walker} S.~A.,  2021, \mn@doi [\mnras]
  {10.1093/mnras/stab1768}, \href
  {https://ui.adsabs.harvard.edu/abs/2021MNRAS.506..139M} {506, 139}

\bibitem[\protect\citeauthoryear{{Miyaji} et~al.,}{{Miyaji}
  et~al.}{2015}]{Miyaji2015}
{Miyaji} T.,  et~al., 2015, \mn@doi [\apj] {10.1088/0004-637X/804/2/104}, \href
  {https://ui.adsabs.harvard.edu/abs/2015ApJ...804..104M} {804, 104}

\bibitem[\protect\citeauthoryear{{Nordlund} \& {Padoan}}{{Nordlund} \&
  {Padoan}}{1999}]{Nordlund1999}
{Nordlund} {\r{A}}.~K.,  {Padoan} P.,  1999, in {Franco} J.,  {Carraminana} A.,
   eds, Interstellar Turbulence. p.~218 (\mn@eprint {arXiv} {astro-ph/9810074})

\bibitem[\protect\citeauthoryear{{Passot} \& {V{\'a}zquez-Semadeni}}{{Passot}
  \& {V{\'a}zquez-Semadeni}}{1998}]{Passot1998}
{Passot} T.,  {V{\'a}zquez-Semadeni} E.,  1998, \mn@doi [\pre]
  {10.1103/PhysRevE.58.4501}, \href
  {https://ui.adsabs.harvard.edu/abs/1998PhRvE..58.4501P} {58, 4501}

\bibitem[\protect\citeauthoryear{{Roncarelli}, {Ettori}, {Borgani}, {Dolag},
  {Fabjan}  \& {Moscardini}}{{Roncarelli} et~al.}{2013}]{Roncarelli2013}
{Roncarelli} M.,  {Ettori} S.,  {Borgani} S.,  {Dolag} K.,  {Fabjan} D.,
  {Moscardini} L.,  2013, \mn@doi [\mnras] {10.1093/mnras/stt654}, \href
  {https://ui.adsabs.harvard.edu/abs/2013MNRAS.432.3030R} {432, 3030}

\bibitem[\protect\citeauthoryear{{Sanders} et~al.,}{{Sanders}
  et~al.}{2020}]{Sanders2020}
{Sanders} J.~S.,  et~al., 2020, \mn@doi [\aap] {10.1051/0004-6361/201936468},
  \href {https://ui.adsabs.harvard.edu/abs/2020A&A...633A..42S} {633, A42}

\bibitem[\protect\citeauthoryear{{Simionescu} et~al.,}{{Simionescu}
  et~al.}{2012}]{Simionescu2012}
{Simionescu} A.,  et~al., 2012, \mn@doi [\apj] {10.1088/0004-637X/757/2/182},
  \href {https://ui.adsabs.harvard.edu/abs/2012ApJ...757..182S} {757, 182}

\bibitem[\protect\citeauthoryear{{Simionescu} et~al.,}{{Simionescu}
  et~al.}{2019}]{Simionescu2019}
{Simionescu} A.,  et~al., 2019, \mn@doi [\ssr] {10.1007/s11214-019-0590-1},
  \href {https://ui.adsabs.harvard.edu/abs/2019SSRv..215...24S} {215, 24}

\bibitem[\protect\citeauthoryear{{Urban} et~al.,}{{Urban}
  et~al.}{2014}]{Urban2014}
{Urban} O.,  et~al., 2014, \mn@doi [\mnras] {10.1093/mnras/stt2209}, \href
  {https://ui.adsabs.harvard.edu/abs/2014MNRAS.437.3939U} {437, 3939}

\bibitem[\protect\citeauthoryear{{Wan}, {Mantz}, {Sayers}, {Allen}, {Morris}
  \& {Golwala}}{{Wan} et~al.}{2021}]{Wan2021}
{Wan} J.~T.,  {Mantz} A.~B.,  {Sayers} J.,  {Allen} S.~W.,  {Morris} R.~G.,
  {Golwala} S.~R.,  2021, arXiv e-prints, \href
  {https://ui.adsabs.harvard.edu/abs/2021arXiv210109389W} {p. arXiv:2101.09389}

\bibitem[\protect\citeauthoryear{{XRISM Science Team}}{{XRISM Science
  Team}}{2020}]{XRISM2020}
{XRISM Science Team} 2020, arXiv e-prints, \href
  {https://ui.adsabs.harvard.edu/abs/2020arXiv200304962X} {p. arXiv:2003.04962}

\bibitem[\protect\citeauthoryear{{Zhuravleva}, {Churazov}, {Kravtsov}, {Lau},
  {Nagai}  \& {Sunyaev}}{{Zhuravleva} et~al.}{2013}]{Zhuravleva2013}
{Zhuravleva} I.,  {Churazov} E.,  {Kravtsov} A.,  {Lau} E.~T.,  {Nagai} D.,
  {Sunyaev} R.,  2013, \mn@doi [\mnras] {10.1093/mnras/sts275}, \href
  {https://ui.adsabs.harvard.edu/abs/2013MNRAS.428.3274Z} {428, 3274}

\bibitem[\protect\citeauthoryear{{Zhuravleva} et~al.,}{{Zhuravleva}
  et~al.}{2014}]{Zhuravleva2014}
{Zhuravleva} I.,  et~al., 2014, \mn@doi [\apjl] {10.1088/2041-8205/788/1/L13},
  \href {https://ui.adsabs.harvard.edu/abs/2014ApJ...788L..13Z} {788, L13}

\bibitem[\protect\citeauthoryear{{Zhuravleva} et~al.,}{{Zhuravleva}
  et~al.}{2015}]{Zhuravleva2015}
{Zhuravleva} I.,  et~al., 2015, \mn@doi [\mnras] {10.1093/mnras/stv900}, \href
  {https://ui.adsabs.harvard.edu/abs/2015MNRAS.450.4184Z} {450, 4184}

\bibitem[\protect\citeauthoryear{{Zhuravleva} et~al.,}{{Zhuravleva}
  et~al.}{2016}]{Zhuravleva2016}
{Zhuravleva} I.,  et~al., 2016, \mn@doi [\mnras] {10.1093/mnras/stw520}, \href
  {https://ui.adsabs.harvard.edu/abs/2016MNRAS.458.2902Z} {458, 2902}

\bibitem[\protect\citeauthoryear{{Zhuravleva}, {Allen}, {Mantz}  \&
  {Werner}}{{Zhuravleva} et~al.}{2018}]{Zhuravleva2018}
{Zhuravleva} I.,  {Allen} S.~W.,  {Mantz} A.,   {Werner} N.,  2018, \mn@doi
  [\apj] {10.3847/1538-4357/aadae3}, \href
  {https://ui.adsabs.harvard.edu/abs/2018ApJ...865...53Z} {865, 53}

\bibitem[\protect\citeauthoryear{{Zhuravleva}, {Chen}, {Churazov},
  {Schekochihin}, {Zhang}  \& {Nagai}}{{Zhuravleva}
  et~al.}{2022}]{Zhuravleva2022}
{Zhuravleva} I.,  {Chen} M.~C.,  {Churazov} E.,  {Schekochihin} A.~A.,  {Zhang}
  C.,   {Nagai} D.,  2022, arXiv e-prints, \href
  {https://ui.adsabs.harvard.edu/abs/2022arXiv221011544Z} {p. arXiv:2210.11544}

\makeatother
\end{thebibliography}

\appendix

\section{Temperature profiles}
\label{apx:temp}

\begin{figure*}
\centering
\includegraphics{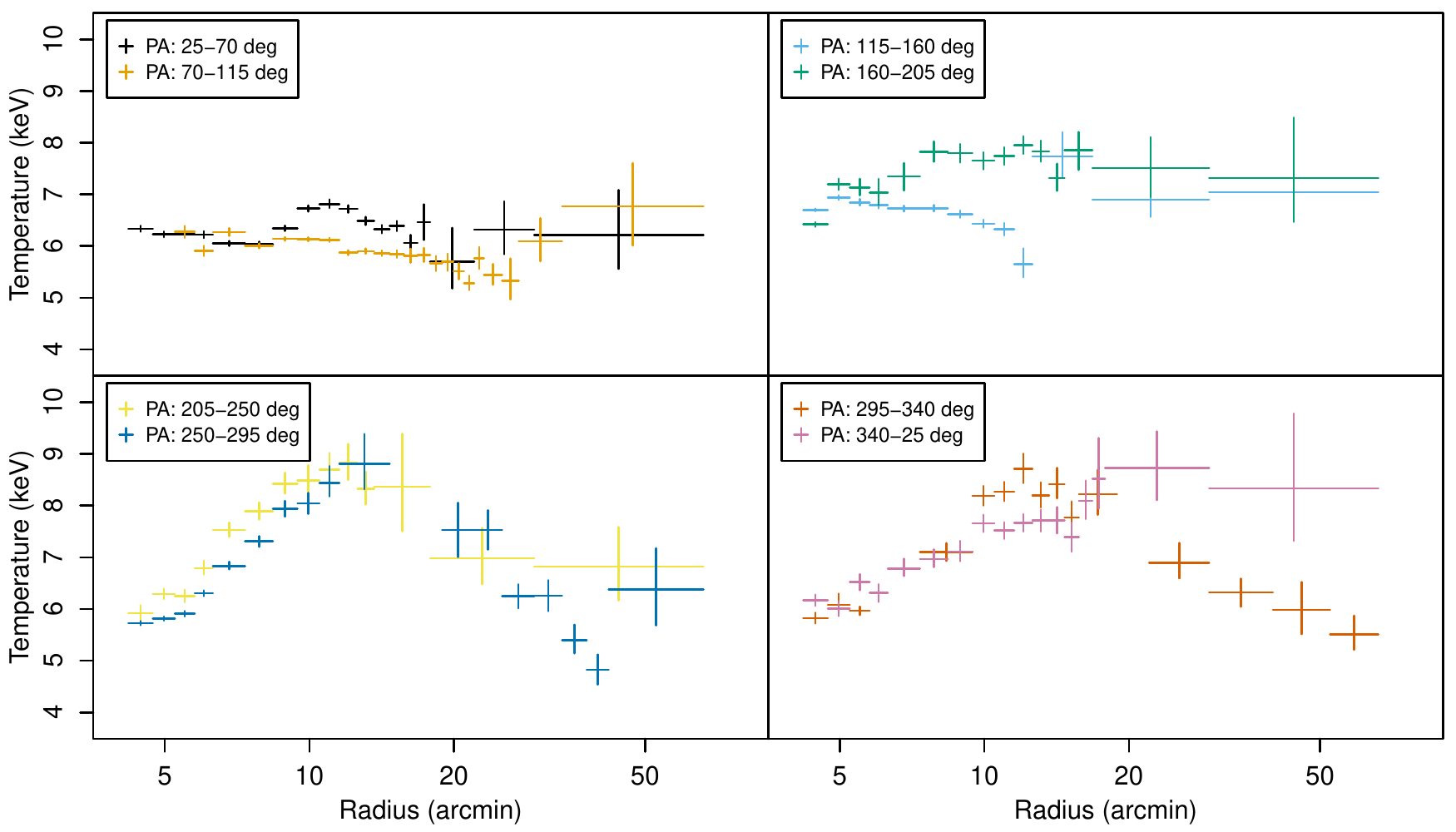} \\
\caption{Projected temperature profiles in 8 different sectors of Perseus}
\label{fig:Tprof}
\end{figure*}

Figure~\ref{fig:Tprof} shows projected temperature profiles for in 8 sectors from the {\it Chandra} data, obtained using the spectral analysis methodology described in \citet{Mantz2014, Mantz2022}.
In particular, the spatial variation of the Galactic equivalent absorbing column density measured by the LAB \ion{H}{I} survey was accounted for, and its overall normalization was simultaneously fitted to the data for each sector along with the temperature as a function of radius.
Unlike the results for Perseus used by \citet{Mantz2022}, here we are neither modeling the total mass profile nor constraining deprojected profiles of temperature and density, but more simply fitting for the projected brightness and temperature in partial annuli, while accounting for the complex absorption across the cluster image.
Metallicities were included as free parameters in the fit, though linked between adjacent radial bins for an overall-courser resolution than the temperature profiles shown.
The sectors used in this analysis are those identified by \citet{Mantz2022}, which differ somewhat from those adopted in this work. To obtain a temperature to compute the sound speed appropriate for a given region in Section \ref{sec:results}, we created linear interpolations of the temperature profiles for each sector shown in Figure \ref{fig:Tprof}. We then evaluated each interpolated function at the appropriate radius, identified the overlap between the sectors in this work and those in \cite{Mantz2022}, and took the weighted average based on overlap in azimuths. The radii of a few shells extend beyond the final data point of the temperature profiles by $\approx 5$ arcmin, yet were still within the total range covered by that data point. In those cases, we simply took the temperature of the last data point as the temperature for the shell. Although this method is somewhat simplistic, we note that the sound speed depends only the square root of the temperature and the temperature varies by only $\approx 4$ keV in the radial range of interest. Therefore, the uncertainties from the MCMC samples of the cluster surface brightness log-normal standard deviation $\sigma_f$ dominate the error in the computed turbulent velocity.

\section{Model visualizations for all shells}
\label{apx:mvis}

Figures \ref{fig:amodelvis} shows the model visualizations for all 28 shells in the Perseus field. 
\begin{figure*}
\centering
\includegraphics[width=1.05\textwidth]{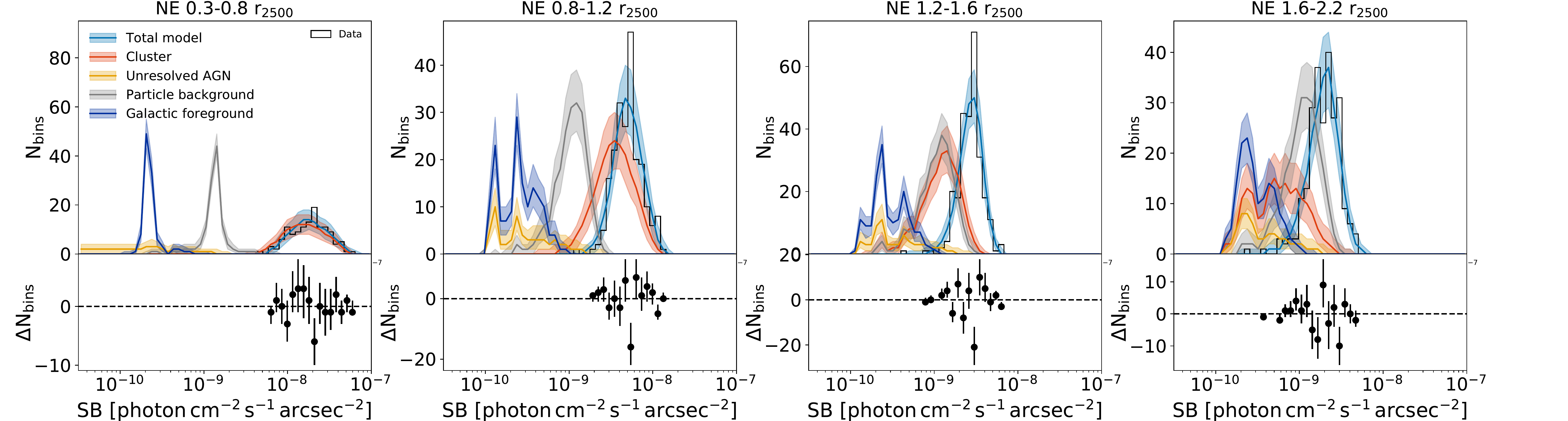} \\
\includegraphics[width=1.05\textwidth]{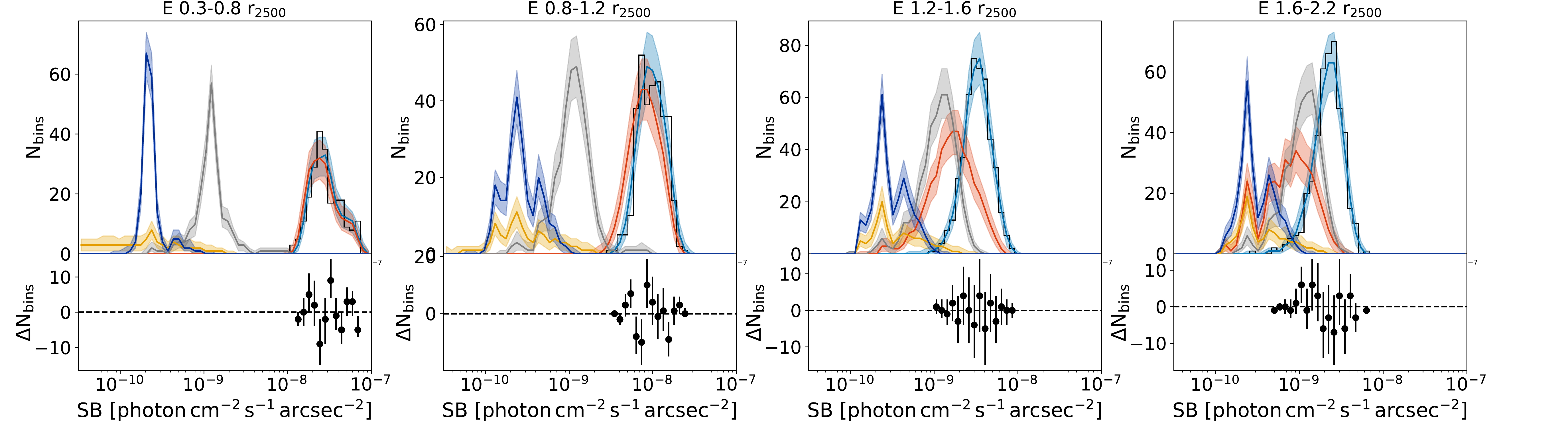} \\
\includegraphics[width=1.05\textwidth]{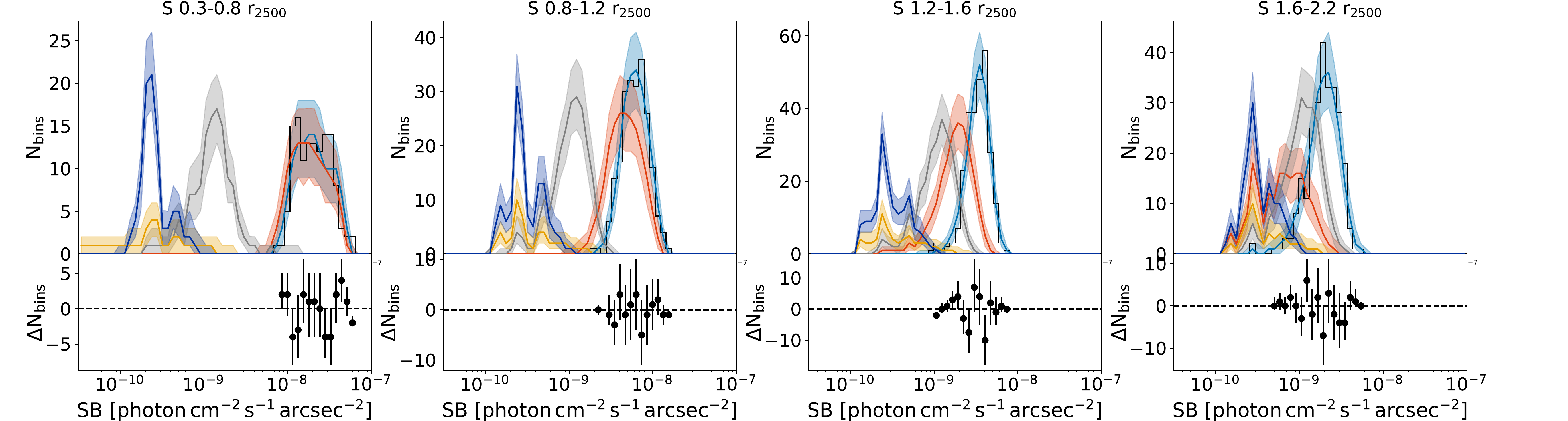} \\
\includegraphics[width=1.05\textwidth]{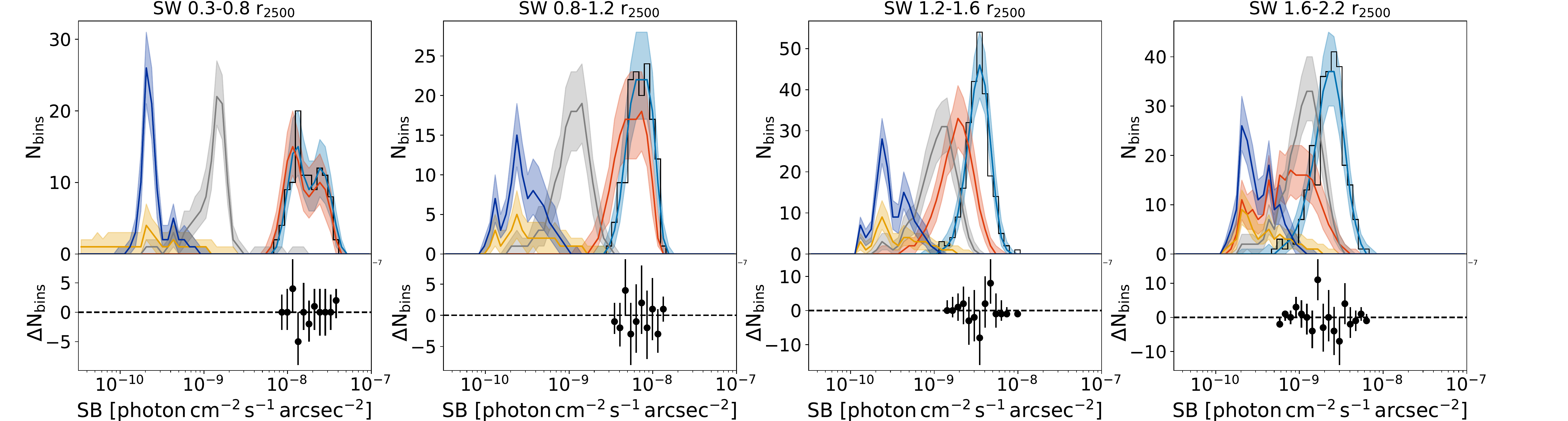} \\
\caption{Forward-simulated model histograms resulting from the MCMC-sampled posterior distributions, and compared with the observed 0.6--3.5\,keV surface brightness data in all 28 shells in the Perseus Cluster. The model error envelopes were created by repeating the forward simulation process 10,000 times, and taking the 14th and 86th percentiles at each histogram bin as the lower and upper model boundaries. We note that the residual plots below each panel are purely meant as a visual aid, as the errors are correlated. }
\label{fig:amodelvis}
\end{figure*}

\begin{figure*}
 \ContinuedFloat 

\centering

\includegraphics[width=1.05\textwidth]{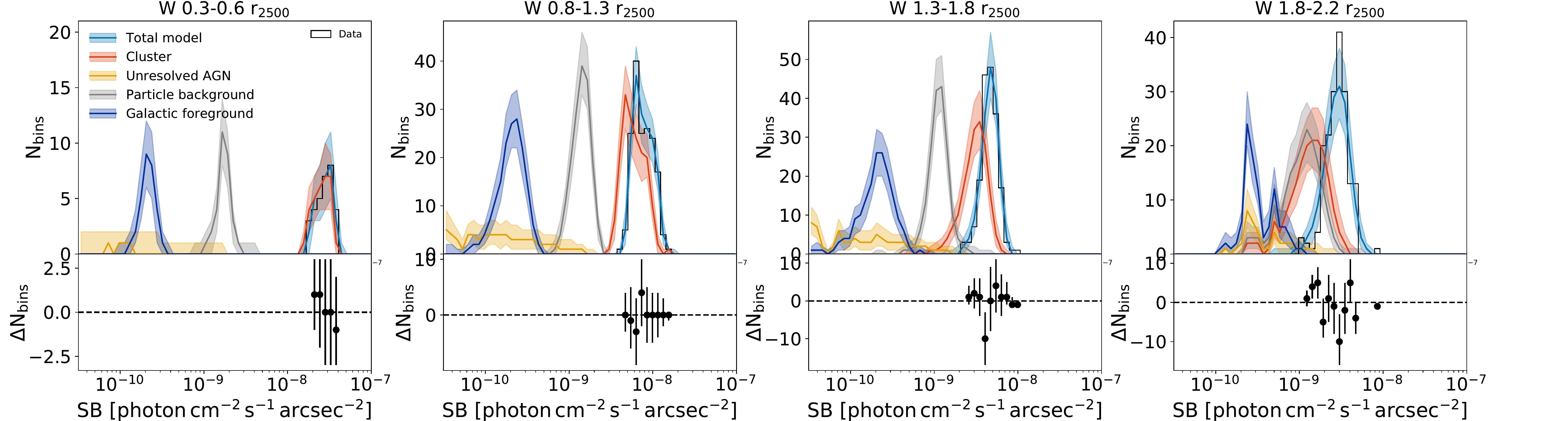} \\
\includegraphics[width=1.05\textwidth]{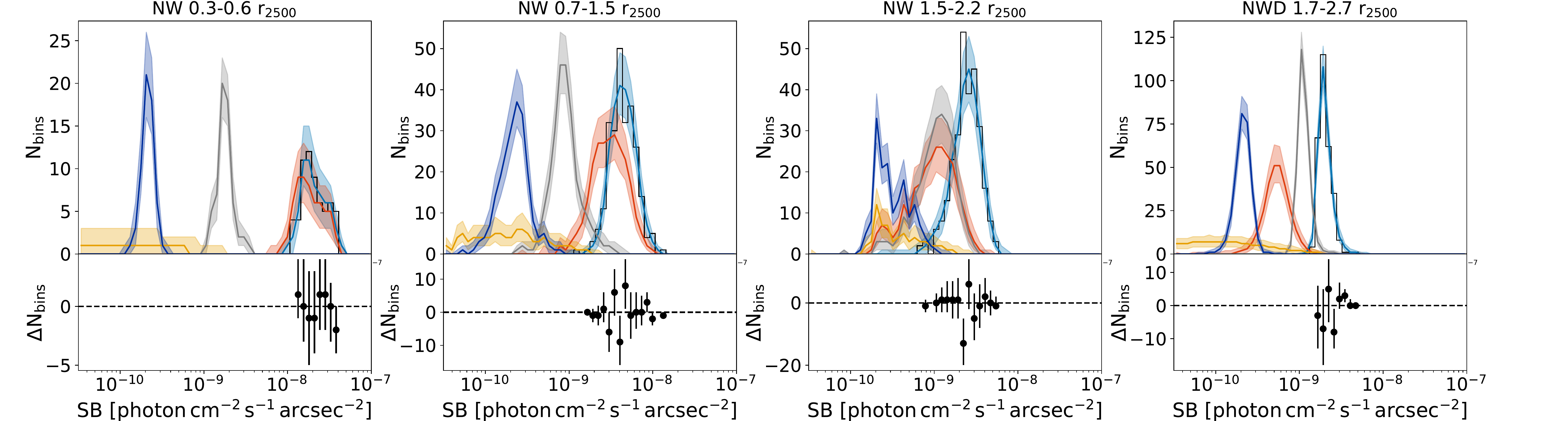} \\
\includegraphics[width=1.05\textwidth]{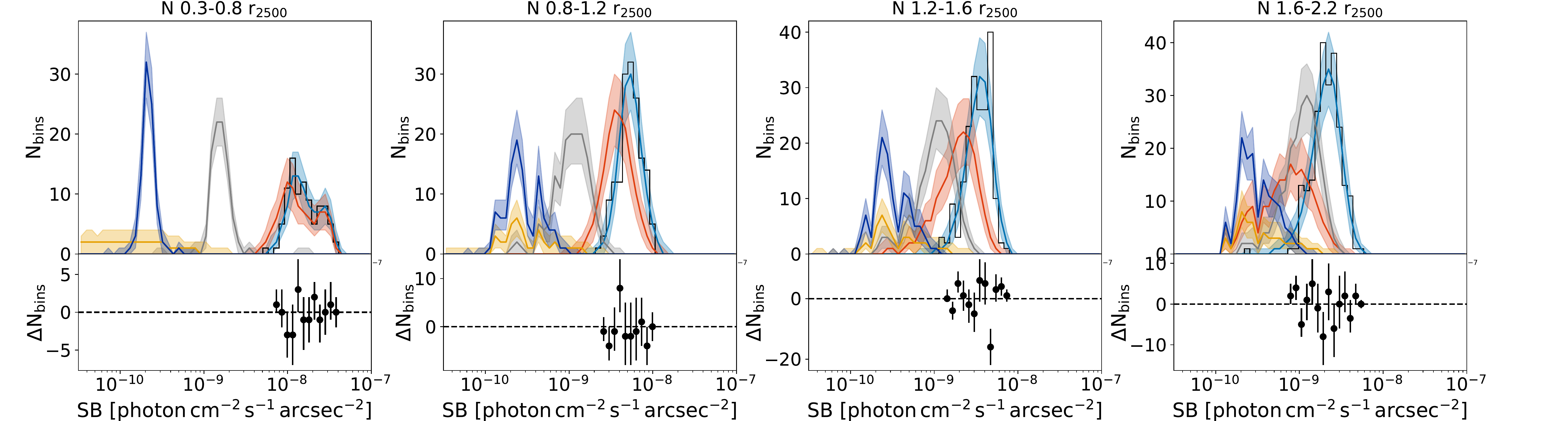} \\
\caption{\textit{(Cont.)} Forward-simulated model histograms resulting from the MCMC-sampled posterior distributions, and compared with the observed 0.6--3.5\,keV surface brightness data in all 28 shells in the Perseus Cluster. The model error envelopes were created by repeating the forward simulation process 10,000 times, and taking the 14th and 86th percentiles at each histogram bin as the lower and upper model boundaries. We note that the residual plots below each panel are purely meant as a visual aid, as the errors are correlated. }
\end{figure*}

\bsp	
\label{lastpage}
\end{document}